\documentclass[12pt,a4paper]{article}

\usepackage{epsf}
\usepackage{latexsym,amssymb,euscript}
\usepackage[dvips]{graphicx}
\usepackage[numbers,sort&compress]{natbib}
\usepackage{amsmath}
\usepackage{slashed}
\usepackage{booktabs}
\usepackage{hyperref}
\usepackage{braket}
\usepackage{chngcntr}
\usepackage{bbold}
\usepackage{graphics}
\usepackage{graphicx}
\usepackage{caption}
\usepackage{subcaption}
\usepackage{pdfpages}
\usepackage[titletoc]{appendix}
\graphicspath{{./figures/}}
\hypersetup{
 linktocpage = true,
 linkcolor = blue,
 urlcolor = blue,
 colorlinks = true,
 linkcolor = teal,
 anchorcolor = blue,
 citecolor = teal,
 urlcolor = teal,
 pdfstartview = {XYZ null null 1.25} 
           }
\usepackage[left=3cm, right=2cm]{geometry}
\usepackage{pstricks}
\usepackage{color}
\usepackage{float}

\newcommand{\tarr}{
\begin{array}}
\newcommand{\earr}{\end{array}}
\newcommand*\rfrac[2]{{}^{#1}\!/_{#2}} 

\begin{document}

\begin{titlepage}

\begin{center}
  {\LARGE \bf High-energy resummation in heavy-quark pair hadroproduction}
\end{center}

\vskip 0.5cm

\centerline{
A.D.~Bolognino$^{1,2,*}$,
F.G.~Celiberto$^{3,4\dagger}$,
M.~Fucilla$^{1\ddagger}$,
D.Yu.~Ivanov$^{5,6\S}$, 
and A.~Papa$^{1,2\P}$}

\vskip .6cm

\centerline{${}^1$ {\sl Dipartimento di Fisica, Universit\`a della Calabria}}
\centerline{\sl I-87036 Arcavacata di Rende, Cosenza, Italy}
\vskip .2cm
\centerline{${}^2$ {\sl Istituto Nazionale di Fisica Nucleare, Gruppo collegato
      di Cosenza}}
\centerline{\sl I-87036 Arcavacata di Rende, Cosenza, Italy}
\vskip .2cm
\centerline{${}^3$ {\sl Dipartimento di Fisica, Universit\`a degli Studi di Pavia, I-27100 Pavia, Italy}}
\vskip .2cm
\centerline{${}^4$ {\sl INFN, Sezione di Pavia, I-27100 Pavia, Italy}}
\vskip .2cm
\centerline{${}^5$ {\sl Sobolev Institute of Mathematics, 630090 Novosibirsk,
    Russia}}
\vskip .2cm
\centerline{${}^6$ {\sl Novosibirsk State University, 630090 Novosibirsk,
    Russia}}
\vskip 2cm

\begin{abstract}
The inclusive hadroproduction of two heavy quarks, featuring a large separation
in rapidity, is proposed as a novel probe channel of the
Balitsky-Fadin-Kuraev-Lipatov (BFKL) approach.
In a theoretical setup which includes full resummation of leading logarithms
in the center-of-mass energy and partial resummation of the next-to-leading
ones, predictions for the cross section and azimuthal coefficients are presented
for kinematic configurations typical of current and possible future
experimental analyses at the LHC. 
\end{abstract}

\vskip .5cm

$^{*}${\it e-mail}:
ad.bolognino@unical.it

$^{\dagger}${\it e-mail}:
francescogiovanni.celiberto@unipv.it

$^{\ddagger}${\it e-mail}:
mike.fucilla@libero.it

$^{\S}${\it e-mail}:
d-ivanov@math.nsc.ru

$^{\P}${\it e-mail}:
alessandro.papa@fis.unical.it

\end{titlepage}


\section{Introduction}
\label{introduction}

The study of high-energy reactions falling in the so-called \emph{semi-hard}
sector~\cite{Gribov:1984tu}, where the scale hierarchy, $s \gg Q^2 \gg
\Lambda_{\rm QCD}^2$ ($s$ is the squared center-of-mass energy, $Q$ the hard
scale given by the process kinematics and $\Lambda_{\rm QCD}$ the QCD mass scale),
strictly holds, definitely represents an excellent channel to probe and deepen
our knowledge of strong interactions in kinematic ranges so far unexplored.

In the Regge limit, $s \gg|t|$, fixed-order calculations in perturbative QCD
miss the effect of large energy logarithms, entering the perturbative series
with a power increasing along with the order, thus compensating the smallness
of the strong coupling, $\alpha_s$. The Balitsky-Fadin-Kuraev-Lipatov
(BFKL)~\cite{BFKL} approach represents the most powerful tool to resum to
all orders, both in the leading (LLA) and the next-to-leading (NLA)
approximation, these large-energy logarithmic contributions. In the BFKL
framework, the cross section of hadronic processes can be expressed as the
convolution of two impact factors, related to the transition from each
colliding particle to the respective final-state object, and a
process-independent Green's function. The evolution of the BFKL Green's
function is controlled by an integral equation, whose kernel is known at the
next-to-leading order (NLO) both for forward scattering ({\it i.e.} for $t=0$
and color singlet in the $t$-channel)~\cite{Fadin:1998py,Ciafaloni:1998gs} and
for any fixed, not growing with $s$, momentum transfer $t$ and any possible
two-gluon color state in the $t$-channel~\cite{Fadin:1998jv,FG00,FF05}.

Our ability to study reactions in the BFKL approach is however restricted by
the exiguous number of available impact factors, since just few of them are
known with NLO accuracy: 1) colliding-parton (quarks and gluons) impact
factors~\cite{fading,fadinq,Cia,Ciafaloni:2000sq}, which represent the common
basis for the calculation of the 2) forward-jet impact
factor~\cite{bar1,bar2,Caporale:2011cc,Ivanov:2012ms,Colferai:2015zfa} and of
the 3) forward light-charged hadron one~\cite{Ivanov:2012iv}, 4) the impact
factor describing the $\gamma^*$ to light-vector-meson leading twist
transition~\cite{IKP04}, and 5) the $\gamma^*$ to $\gamma^*$
transition~\cite{gammaIF,Balitsky2012}.

Pursuing the goal to get a more exhaustive comprehension of this high-energy
regime, a significant range of semi-hard reactions (see
Ref.~\cite{Celiberto:2017ius} for applications) has been proposed so far:
the diffractive leptoproduction of
one~\cite{Bolognino:2018rhb,Bolognino:2018mlw,Bolognino:2019bko} or two light
vector mesons~\cite{Ivanov:2004pp,Ivanov:2005gn,Ivanov:2006gt,Enberg:2005eq},
the inclusive hadroproduction of two jets featuring large transverse momenta
and well separated in rapidity (Mueller--Navelet channel~\cite{Mueller:1986ey}),
for which several phenomenological studies have appeared so far~\cite{Colferai:2010wu,Caporale:2012ih,Ducloue:2013wmi,Ducloue:2013bva,Caporale:2013uva,Ducloue:2014koa,Caporale:2014gpa,Ducloue:2015jba,Caporale:2015uva,Celiberto:2015yba,Celiberto:2015mpa,Celiberto:2016ygs,Celiberto:2016vva,Caporale:2018qnm,Chachamis:2015crx}, the inclusive detection of two light-charged rapidity-separated
hadrons~\cite{Celiberto:2016hae,Celiberto:2016zgb,Celiberto:2017ptm}, three-
and four-jet hadroproduction~\cite{Caporale:2015vya,Caporale:2015int,Caporale:2016soq,Caporale:2016vxt,Caporale:2016xku,Celiberto:2016vhn,Caporale:2016djm,Caporale:2016lnh,Caporale:2016zkc}, $J/\Psi$-jet~\cite{Boussarie:2017oae},
hadron-jet~\cite{Bolognino:2018oth,Bolognino:2019yqj,Bolognino:2019cac} and
forward Drell--Yan dilepton production~\cite{Motyka:2014lya,Brzeminski:2016lwh,Celiberto:2018muu} with a possible backward-jet
tag~\cite{Golec-Biernat:2018kem,Deak:2018obv}.

\begin{figure}
\centering
\includegraphics[width=0.5\textwidth]{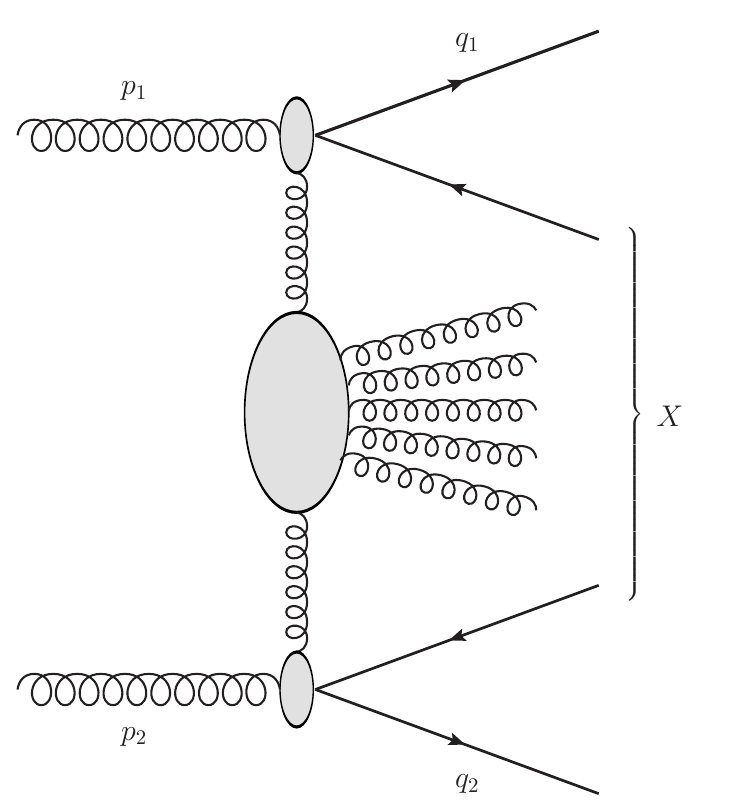}
\caption{Diagrammatic representation of the heavy-quark pair hadroproduction in
  the case when a heavy quark with transverse momentum $q_1(q_2)$ from the
  upper (lower) vertex is tagged.}
\label{fig:hadroproduction}
\end{figure}

In this work we introduce and study within NLA BFKL accuracy a novel semi-hard
reaction, \emph{i.e.} the inclusive emission of two rapidity-separated
heavy quarks in the collision of two protons ({\em hadroproduction}).
In Refs.~\cite{Celiberto:2017nyx,Bolognino:2019ouc} a process with the same
final state was considered, but produced through the collision of two
(quasi-)real photons ({\em photoproduction}) emitted by two interacting
electron and positron beams according to the equivalent-photon
approximation (EPA). For center of mass energies much larger than the hard
scale of the process, given here by the heavy-quark mass, the prerequisites
are fulfilled for a theoretical description within the BFKL approach.
Similarly to the treatment of the photoproduction case in
Refs.~\cite{Celiberto:2017nyx,Bolognino:2019ouc}, we will convolute
leading-order impact factors with the NLA BFKL Green's function. In this
approximation, the hadroproduction process is initiated at partonic level
by a gluon-gluon collision:
\begin{equation}
\label{process}
g(p_1) \ + \ g(p_2) \ \to \ Q\text{-jet}(q_1) \ + \ X \ + \ Q\text{-jet}(q_2)
\;,
\end{equation}
where $Q$ stands for a charm/bottom quark {\em or} the respective antiquark.
In Fig.~\ref{fig:hadroproduction} we present a pictorial description of this
process, in the case when the tagged object from the upper (lower) vertex is a
heavy quark with transverse momentum $q_1$ ($q_2$).

The aim of this paper is to provide predictions for cross section and
azimuthal coefficients of the expansion in the (cosine of the) relative
angle in the transverse plane between the flight directions of the two
tagged heavy quarks, to be compared with current and future experimental
analyses at the LHC. We will see that in the same kinematical conditions
where the photoproduction process was considered in
Refs.~\cite{Celiberto:2017nyx,Bolognino:2019ouc}, the hadroproduction mechanism
leads to a much higher cross section. Moreover, we will consider in detail
the inclusive hadroproduction of two bottom quarks and present a
phenomenological analysis tailored on the kinematics and the energies of the
LHC, proposing it as a new channel for the investigation of the BFKL dynamics
at hadron colliders.

The work is organized as follows: 
Section~\ref{theory} is to set the theoretical framework up;
Section~\ref{numerical_analysis} is devoted to our results for cross sections
and azimuthal coefficients and correlations as a function of the rapidity
interval, $\Delta Y$, between the tagged heavy quarks;
Section~\ref{summary_outlook} carries our closing statements and some outlook.

\section{Theoretical setup}
\label{theory}

\begin{figure}
\centering
\includegraphics[width=0.5\textwidth]{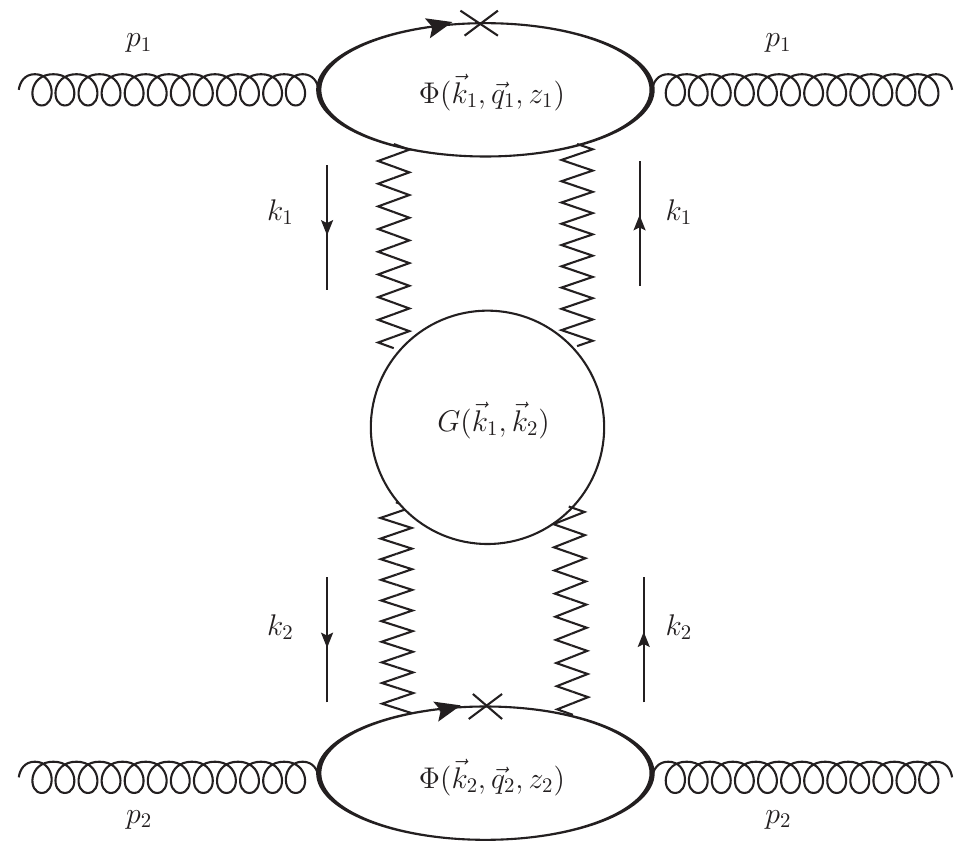}
\caption{Schematic representation of the BFKL factorization for the heavy-quark pair hadroproduction.}
\label{fig:BFKL_factorization}
\end{figure}

For the process under consideration (see Fig.~\ref{fig:hadroproduction}) we plan
to construct the cross section, differential in some of the kinematic variables
of the tagged heavy quark or antiquark, and some azimuthal correlations between
the tagged fermions. In the BFKL approach the cross section takes the
factorized form, diagrammatically represented in
Fig.~\ref{fig:BFKL_factorization}, given by the convolution of the impact
factors for the transition from a real gluon to a heavy quark-antiquark pair
with the BFKL Green's function $G$.

In our calculation we will partially include NLA resummation effects, by
taking the BFKL Green's function in the NLA, while the impact factors are kept
at leading order.   

\subsection{Impact factor}
\label{impact_factor}

The (differential) impact factor for the hadroproduction of a heavy-quark pair
reads~\footnote{See \hyperlink{app:IF_def-link}{Appendix~A} for a sketch of its calculation.}
\begin{equation}
\begin{split}
d\Phi^{\lbrace{Q\bar{Q}\rbrace}}_{gg}(\vec{k},\vec{q},z)& =\frac{\alpha_s^2
\sqrt{N_c^2-1}}{2\pi N_c}\left[\left(m^2\left(R+\bar{R}\right)^2
+\left(z^2+\bar{z}^2 \right)\left(\vec{P}+\vec{\bar{P}}\right)^2\right)
\right. \\ & \left. -\frac{N_c^2}{N_c^2-1}\left(2m^2R\bar{R}
+\left(z^2+\bar{z}^2\right)2\vec{P} \cdot \vec{\bar{P}}\right)\right]
\; d^2\vec{q} \; dz\;,
\end{split}
\label{eq:imp.fac2}
\end{equation}
where $R$, $\bar{R}$, $\vec{P}$ and $\vec{\bar{P}}$ are defined as 
\begin{equation}
\label{laR}
R = \frac{1}{m^2+\vec{q}^{\;2}}-\frac{1}{m^2+(\vec{q}-z\vec{k})^{2}} \;,  
\end{equation}
\begin{equation}
\label{LaRbar}
\bar{R} = \frac{1}{m^2+(\vec{q}-z\vec{k})^{2}}-\frac{1}{m^2+(\vec{q}-\vec{k})^{2}} \;, 
\end{equation}
\begin{equation}
\label{laP}
\vec{P} = \frac{\vec{q}}{m^2+\vec{q}^{\;2}}-\frac{\vec{q}-z\vec{k}}{m^2+(\vec{q}
  -z\vec{k})^{2}} \;,  
\end{equation}
\begin{equation}
\label{LaPbar}
  \vec{\bar{P}} = \frac{\vec{q}-z\vec{k}}{m^2+(\vec{q}-z\vec{k})^{2}}
  -\frac{\vec{q}-\vec{k}}{m^2+(\vec{q}-\vec{k})^{2}} \;.  
\end{equation}
Here $\alpha_s$ denotes the QCD coupling, $N_c$ gives the number of colors, $m$
stands for the heavy-quark mass, $z$ and $\bar{z} \equiv 1-z$ are the
longitudinal momentum fractions of the quark and antiquark produced in the
same vertex and $k$, $q$, $k-q$ represent the transverse momenta with respect
to the gluons collision axis of the Reggeized gluon, the produced quark and
antiquark, respectively.

In the following we will need the projection of the impact factors onto the
eigenfunctions of the leading-order BFKL kernel, to get their so called
$(n, \nu)$-representation. We get 
\[
\frac{d\Phi_{gg}^{\lbrace{Q\bar{Q}\rbrace}}\left(n,\nu,\vec{q},z\right)}{d^2\vec{q}
  \; dz} \equiv \int\frac{d^2\vec{k}}{\pi\sqrt{2}}(\vec{k}^{\;2})^{i\nu-\frac{3}{2}}
e^{in\theta}\frac{d\Phi^{\lbrace{Q\bar{Q}\rbrace}}_{gg}(\vec{k},\vec{q},z)}{d^2\vec{q}
  \; dz} 
\]
\[
= \frac{\alpha_s^2 \sqrt{N_c^2-1}}{2\pi N_c}\left\{ m^2 \left( I_3
- 2\frac{I_2(0)}{m^2+\vec{q}^{\;2}} \right) + (z^2 + \bar{z}^2)
\left( -m^2 \left(I_3 - 2\frac{I_2(0)}{m^2+\vec{q}^{\;2}} \right)
+ \frac{I_2(1)}{m^2 + \vec{q}^{\; 2}} \right) \right.
\]
\[
\left. - \frac{N_c^2}{N_c^2-1} \Bigg[ 2 m^2 \left[\left( z^2 + \bar{z}^2
 - 1 \right) \left( 1 -  \left(z^2\right)^{\frac{1}{2}-i\nu} \right) \right]
 \frac{I_2(0)}{m^2+ \vec{q}^{\; 2}} + \left[ 2m^2(z^2 + \bar{z}^2 -1)
\left(z^2\right)^{\frac{1}{2}-i\nu} \right] 
\right.
\]
\[
\left.
\left( I_3 - \frac{I_4(0)}{\left(z^2\right)^{\frac{1}{2}-i\nu}} \right)
- (z^2 + \bar{z}^2) \bigg[ (1-z)^2 I_4(1) - \frac{\left( 1
    - \left(z^2\right)^{\frac{1}{2}-i\nu} \right)}{m^2 + \vec{q}^{\;2}} I_2(1) \bigg]
\Bigg] \right\} 
\]
\begin{equation}
\equiv \alpha_s^2 \; e^{in \varphi} c(n, \nu, \vec{q},z) \;,
\label{eq:imp.fac projected1}
\end{equation}
where $I_2(\lambda)$, $I_3$ and $I_4(\lambda)$ read
\begin{equation}
\label{AAA}
\begin{split}
I_2\left(\lambda\right)&=\frac{\left(\vec{q}^{\; 2}\right)^{\frac{n}{2}}e^{in\varphi}}
{\sqrt{2}}\frac{1}{\left(m^2+\vec{q}^{\;2}\right)^{\frac{3}{2}+\frac{n}{2}-i\nu-\lambda}}
\frac{\Gamma\left(\frac{1}{2}+\frac{n}{2}+i\nu+\lambda\right)
  \Gamma\left(\frac{1}{2}+\frac{n}{2}-i\nu-\lambda\right)}{\Gamma\left(1+n\right)}\\ &\times\frac{\left(\frac{1}{2}+\frac{n}{2}-i\nu-\lambda\right)}
     {\left(-\frac{1}{2}+\frac{n}{2}+i\nu+\lambda\right)}\;
     _2F_1\left(-\frac{1}{2}+\frac{n}{2}+i\nu+\lambda,\frac{3}{2}+\frac{n}{2}
     -i\nu-\lambda,1+n,\zeta\right) \;,
\end{split} 
\end{equation} 
\vspace{0.5 cm}
\begin{equation}
\label{I3_final}
\begin{split}
  I_3= \frac{\left(\vec{q}^{\; 2}\right)^{\frac{n}{2}}e^{in\varphi}}{\sqrt{2}}
  & \frac{1}{\left(m^2+\vec{q}^{\;2}\right)^{\frac{5}{2}+\frac{n}{2}-i\nu}}
  \frac{\Gamma\left(\frac{1}{2}+\frac{n}{2}+i\nu\right)
    \Gamma\left(\frac{1}{2}+\frac{n}{2}-i\nu\right)}
       {\Gamma\left(1+n\right)}\frac{\left(\frac{1}{2}+\frac{n}{2}-i\nu\right)}
       {\left(-\frac{1}{2}+\frac{n}{2}+i\nu\right)}\;
       \\ & \times \left(\frac{3}{2}+\frac{n}{2}-i\nu\right)\;
       _2F_1\left(-\frac{1}{2}+\frac{n}{2}+i\nu,\frac{5}{2}+\frac{n}{2}
       -i\nu,1+n,\zeta \right) \;,
\end{split} 
\end{equation}
\vspace{0.5 cm}
\begin{equation}
\begin{split}
  I_4\left(\lambda\right) = & \frac{\left(\vec{q}^{\; 2}\right)^{\frac{n}{2}}
    e^{in\varphi}}{z^2\sqrt{2}}  \frac{\left(\frac{3}{2}-i\nu-\lambda
+\frac{n}{2}\right)}{\left(m^2+\vec{q}^{\; 2}\right)^{\frac{5}{2}-i\nu-\lambda+\frac{n}{2}}} \frac{\Gamma\left(\frac{1}{2}+\frac{n}{2}+i\nu+\lambda \right)\Gamma\left(\frac{1}{2}+\frac{n}{2}-i\nu-\lambda \right)}{\Gamma\left(1+n\right)} \\ & \times \frac{\left(\frac{1}{2}+\frac{n}{2}-i\nu-\lambda \right)}{\left(-\frac{1}{2}+\frac{n}{2}+i\nu+\lambda \right)} \int_0^1 d\Delta \left(1+\frac{\Delta}{z}-\Delta\right)^n \left(1+\frac{\Delta}{z^2}-\Delta\right)^{-\frac{5}{2}+i\nu+\lambda-\frac{n}{2}} \; \\ & \times \; _2F_1\left(-\frac{1}{2}+i\nu+\lambda+\frac{n}{2},\frac{5}{2}-i\nu-\lambda+\frac{n}{2},1+n,\zeta \;  \frac{\left(1+\frac{\Delta}{z}-\Delta\right)^2}{\left(1+\frac{\Delta}{z^2}-\Delta\right)}\right) \;,
\end{split}
\label{I4}
\end{equation}
and $\zeta \equiv \frac{\vec{q}^{\; 2}}{m^2+\vec{q}^{\; 2}}$; the azimuthal
angles $\theta$ and $\varphi$ are defined as $\cos \theta \equiv k_x/|\vec{k}|$
and $\cos \varphi \equiv q_x/|\vec{q}|$. We refer the reader to the
\hyperlink{app:IF_calc-link}{Appendix~B} for details on the derivation of these
results.

\subsection{Kinematics of the process}
\label{kinematics}

For the tagged quark momenta we introduce the standard Sudakov decomposition,
using as light-cone basis the momenta $p_1$ and $p_2$ of the colliding gluons,
\begin{equation}
q = zp_1 + \frac{m^2+ \vec{q}^{\; 2}}{zW^2}p_2+q_{\perp} \;,
\end{equation}
with $W^2=(p_1+p_2)^2=2p_1 \cdot p_2=4E_{g_1}E_{g_2}$ ; $p_1=E_{g_1}(1,\vec{0},1)$
and $p_2=E_{g_2}(1,\vec{0},-1)$, so that
\begin{equation}
\label{2q1p2}
2q \cdot p_2=2 z p_1 \cdot p_2= z W^2 = 2E_{g_2} \left(E + q_{\|}\right) \;,
\end{equation}
\begin{equation}
\label{2q1p1}
2q \cdot p_1 = \frac{m^2+\vec{q}^{\; 2}}{z} = 2E_{g_1} \left(E - q_{\|}\right) \;;
\end{equation}
here $q=(E,\vec{q},q_{\|})$ and the rapidity can be expressed as
\begin{equation}
  y=\frac{1}{2}\ln\frac{\left(E+q_{\|}\right)}{\left(E-q_{ \|}\right)}
  = \ln\left[\frac{2z E_{g_1}} {\sqrt{m^2+\vec{q}^{\; 2}}}\right] \;.
\end{equation}
Accordingly, the rapidities of the two tagged quarks in our process are 
\begin{equation}
\label{rapidity1}
y_1=\ln\left[\frac{2z_1 E_{g_1}} {\sqrt{m^2+\vec{q}_1^{\;2}}}\right]
\hspace{0.2 cm} {\rm{and}} \hspace{0.2 cm} y_2= -\ln\left[\frac{2z_2 E_{g_2}}
  {\sqrt{m^2+\vec{q}_2^{\;2}}}\right]\;,
\end{equation}
whence their rapidity difference is 
\begin{equation}
  \Delta Y \equiv y_1-y_2= \ln \frac{W^2 z_1 z_2}{\sqrt{\left(m^2+\vec{q}_1^{\; 2}
      \right)\left(m^2+\vec{q}_2^{\; 2}\right)}} \;.
\end{equation}
For the semi-hard kinematics we have the requirement 
\begin{equation}
\frac{W^2}{\sqrt{\left(m^2+\vec{q}_1^{\; 2}\right)\left(m^2+\vec{q}_2^{\; 2}\right)}}=\frac{e^{\Delta Y}}{z_1 z_2} \gg 1 \;,
\end{equation}
therefore we will consider the kinematics when $\Delta Y \geq \Delta_0 \sim 1
\div 2$.

In what follows, we will need a cross section differential in the rapidities of
the tagged quarks. For this reason we adopt the change of variables:
\begin{equation*}
z_1 \to y_1=\ln\left[\frac{2z_1 E_{g_1}} {\sqrt{m^2+\vec{q}_1^{\; 2}}}\right]\;, \quad  dy_1=\frac{dz_1}{z_1}\;,
\end{equation*}
\begin{equation*}
z_2 \to y_2=-\ln\left[\frac{2z_2 E_{g_2}} {\sqrt{m^2+\vec{q}_2^{\; 2}}}\right]\;, \quad  dy_2=-\frac{dz_2}{z_2}\;,
\end{equation*}
which implies
\begin{equation*}
dz_1 dz_2=\frac{e^{\Delta Y}\sqrt{m^2+\vec{q}_1^{\; 2}}\sqrt{m^2+\vec{q}_2^{\; 2}}}{W^2} dy_1dy_2 \;.
\end{equation*}

\subsection{The BFKL cross section and azimuthal coefficients} 
\label{CROX}

The differential cross section for the inclusive production of a pair of heavy
quarks separated in rapidity can be cast in the form:
\begin{equation}
\label{crosfin}
\frac{d\sigma_{gg}}{dy_1dy_2d|\vec{q}_1|d|\vec{q}_2|d\varphi_1 d\varphi_2}=\frac{1}{(2\pi)^2} \left[\mathcal{C}_0+2 \sum_{n=1}^{\infty} \cos(n\varphi) \mathcal{C}_n \right],
\end{equation}
where $\varphi=\varphi_1-\varphi_2-\pi$, while $\mathcal{C}_0$ gives the,
$\varphi$-averaged, cross section summed over the azimuthal angles,
$\varphi_{1,2}$, of the produced quarks, and the other coefficients,
$\mathcal{C}_n$, determine the distribution of the relative azimuthal angle
between the two quarks.

The expression for the $\mathcal{C}_n$ coefficient is the following
(see, {\it e.g.}, Ref.~\cite{Caporale:2015uva}):
\begin{equation}
\begin{split}
\label{Cn}
\mathcal{C}_n =& \frac{|\vec{q}_1||\vec{q}_2|\sqrt{m^2+\vec{q}_1^{\; 2}}\sqrt{m_2^{2}+\vec{q}_2^{\; 2}}}{W^2} e^{\Delta Y} \\ & \times \int_{- \infty}^{+ \infty} d\nu \left(\frac{W^2}{s_0}\right)^{\bar{\alpha}_s\left(\mu_R\right)\chi\left(n,\nu\right)+\bar{\alpha}_s^2\left(\mu_R\right)\left(\bar{\chi}\left(n,\nu \right)+\frac{\beta_0}{8N_c}\chi\left(n,\nu\right)\left(-\chi \left(n,\nu\right)+\frac{10}{3}+2\ln\frac{\mu_R^2}{\sqrt{s_1s_2}} \right)\right)} \\ & \times \alpha_s^4 \left(\mu_R\right) c_1\left(n,\nu,\vec{q}_1^{\; 2},z_1 \right)c_2\left(n,\nu,\vec{q}_2^{\; 2},z_2 \right) \left \{ 1+\bar{\alpha}_s \left(\mu_R\right) \left(\frac{\bar{c}_1^{(1)}}{c_1}+\frac{\bar{c}_2^{(1)}}{c_2}\right) \right.\\ & \left. + \bar{\alpha}_s \left(\mu_R\right) \frac{\beta_0}{2N_c} \left(\frac{5}{3}+\ln \frac{\mu_R^2}{s_1s_2}+f\left(\nu\right) \right) + \bar{\alpha}_s^2 \left(\mu_R\right) \ln \left(\frac{W^2}{s_0}\right) \frac{\beta_0}{4N_c} \chi \left(n, \nu \right) f\left(\nu\right) \right \} \;,
\end{split}
\end{equation} 
where
\begin{equation}
  \chi \left(n,\nu\right) = 2 \psi(1)- \psi \left( \frac{n}{2} + \frac{1}{2}
  + i \nu \right) - \psi \left( \frac{n}{2} + \frac{1}{2} - i \nu \right)
\end{equation} 
are the eigenvalues of the leading-order BFKL kernel, with
$\psi = \Gamma'(x)/\Gamma(x)$, and 
\begin{equation}
\beta_0 = \frac{11}{3} N_c - \frac{2}{3} n_f
\end{equation} 
is the first coefficient of the QCD $\beta$-function, responsible for running-coupling effects. 
The function $f\left(\nu \right)$ is defined by
\begin{equation}
i\frac{d}{d\nu}\ln\frac{c_1}{c_2} = 2 \left[f(\nu)-\ln(\sqrt{s_1s_2}) \right] \; ,
\end{equation}   
with $s_i$, $i=1,2$ the hard scales in our two-tagged-quark process, which are
chosen to be equal to $m_i^2+\vec{q}_i^{\; 2}$, and
\begin{equation}
\begin{split}
c_1 \left(n,\nu, \vec{q}_1^{\; 2},z_1 \right) = \frac{1}{e^{in\varphi_1} \alpha_s^2} \frac{d\Phi_{gg}^{\lbrace{Q\bar{Q}\rbrace}}\left(n,\nu,\vec{q}_1,z_1\right)}{d^2\vec{q}_1 \; dz_1} 
\end{split}
\end{equation}
\begin{equation}
\begin{split}
c_2 \left(n,\nu, \vec{q}_2^{\; 2},z_2 \right) = \frac{1}{e^{-in(\varphi_2+\pi)} \alpha_s^2} \left[\frac{d\Phi_{gg}^{\lbrace{Q\bar{Q}\rbrace}}\left(n,\nu,\vec{q}_2,z_2\right)}{d^2\vec{q}_2 \; dz_2}\right]^{*}
\end{split}
\end{equation}
\begin{equation}
\frac{\bar{c}_1^{(1)}}{c_1}+\frac{\bar{c}_2^{(1)}}{c_2}= \chi \left( n,\nu \right) \ln \frac{s_0}{\sqrt{\left(m_1^2+\vec{q}_1^{\; 2}\right) \left(m_2^{2}+\vec{q}_2^{\; 2}\right)}} \;.
\end{equation}
The presence in the latter formula of the combination $\varphi_2 + \pi$ is due
to the fact that, in the second impact factor, the Reggeon is outgoing instead
of incoming. The scale $s_0$ can be arbitrarily chosen, within NLA accuracy;
in this calculation, the choice $s_0= \sqrt{s_1s_2}$ has been made. It is worth
to remark that Eq.~(\ref{Cn}) is written for the general case when two heavy
quarks of different masses $m_1,m_2$ are detected.

\subsection{The proton-proton cross section}
\label{sigma_pp}

In order to pass from the gluon-initiated process to the one initiated by
proton-proton collisions, we must take into account the distribution of the
gluons inside the two colliding particles,
\begin{equation}
\label{dsigma_pp_conv}
d\sigma_{pp} = f_{g_1}(x_1,\mu_{F_1})f_{g_2}(x_2,\mu_{F_2}) d\sigma_{gg} dx_1 dx_2 \;,
\end{equation}
with $f_{g_i}$, $i=1,2$ being the gluon collinear parton distribution functions and $d\sigma_{gg}$ the cross-section in Eq.~(\ref{crosfin}).
Therefore the final expression for our observable is
\begin{equation}
\label{crosfin}
\frac{d\sigma_{pp}}{d(\Delta Y) d\varphi_1 d\varphi_2}=\frac{1}{(2\pi)^2} \left[C_0+2 \sum_{n=1}^{\infty} \cos(n \varphi) C_n \right] \;,
\end{equation}
where
\begin{equation}
\begin{split}
C_n = & 
\int_{q_{1,{\rm{min}}}}^{q_{1,{\rm{max}}}} d|\vec{q}_1| 
\int_{q_{2,{\rm{min}}}}^{q_{2,{\rm{max}}}} d|\vec{q}_2| 
\int_{y_{1,{\rm{min}}}}^{y_{1,{\rm{max}}}} dy_1
\int_{y_{2,{\rm{min}}}}^{y_{2,{\rm{max}}}} dy_2 \;
\delta(y_1 - y_2 - \Delta Y) \\ & \int_{e^{-(y_{1,{\rm{max}}}-y_1)}}^{1} dx_1 f_{g_1}(x_1,\mu_{F_1}) \int_{e^{-(y_{2,{\rm{max}}}+y_2)}}^{1} dx_2 f_{g_2}(x_2,\mu_{F_2}) \; \mathcal{C}_n
\end{split}
\end{equation}
stands for the $n^{\rm th}$ azimuthal coefficient \emph{integrated} over the
$(\vec{q}_{1,2},y_{1,2})$ phase space and the rapidity separation between the
two tagged quarks is kept fixed to $\Delta Y$.

\subsection{The ``box'' $Q \bar Q$ cross section}

\begin{figure}
\centering
\includegraphics[width=0.7\textwidth]{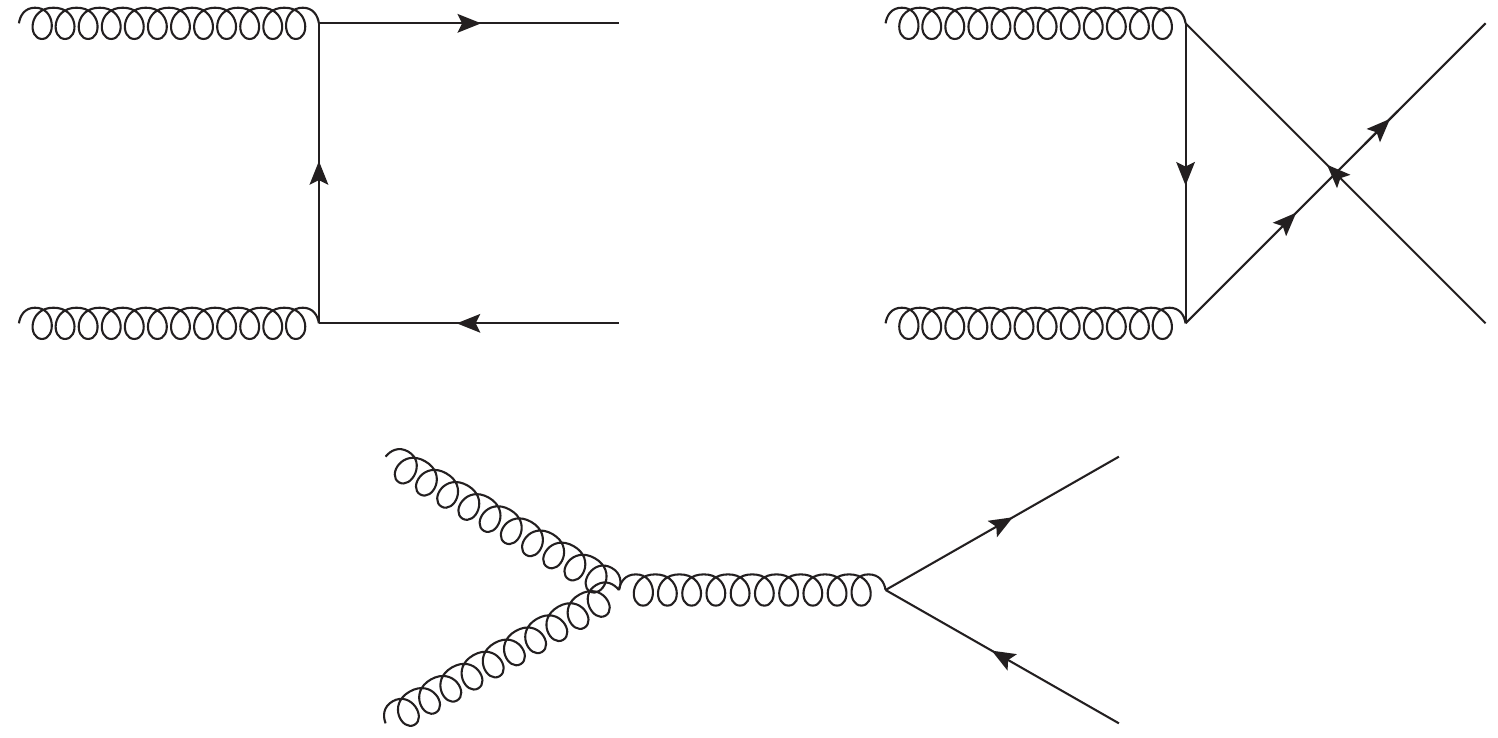}
\caption{Feynman diagrams contributing at the lowest order to the $Q \bar{Q}$
  hadroproduction.}
\label{Box}
\end{figure}

In this section, we consider, for the sake of comparison, the lowest-order QCD
cross section for the production of a heavy quark-antiquark pair in
proton-proton collisions. This process, which we dub ``box'' with a little
abuse of terminology, does not represent a background for the inclusive
reaction of interest in this work when the two detected heavy quarks are of
different flavors or, being of the same flavors, are both quark or both
antiquarks.

The Feynman diagrams contributing to this process at the leading order
are shown in Fig.~\ref{Box}. The differential cross section is presented,
{\it e.g.}, in Ref.~\cite{Ahrens:2010zv} and in our notation takes the form
\begin{equation}
\begin{split}
  \frac{d \sigma_{pp}}{d(\Delta Y)} = \; & \frac{ \pi \alpha_s^2}{s(N_c^2-1)}
  \int_{{\rm min}(q^2_{\rm min},\frac{s}{4\cosh^2(\Delta Y/2)}-m^2)}^{{\rm min}(q^2_{\rm max},
      \frac{s}{4\cosh^2(\Delta Y/2)}-m^2)} \frac{d\vec q^{\: 2}}{M^2}
    \int_{\frac{M^2}{s}}^{1}
    \frac{dx}{x} f_{g_1}(x, \mu_{F_1}) f_{g_2}(M^2/xs, \mu_{F_2}) \\
  & \times \left( C_F \frac{M^4}{t_1 u_1} - C_A \right)
  \left[ \frac{t_1^2+u_1^2}{M^4} + 4 \frac{m^2}{M^2}
    - 4 \frac{m^4}{t_1 u_1} \right] \; ,
\end{split}
\end{equation}
where
\begin{equation}
M^2 = 4 (m^2 + \vec q^{\: 2}) \cosh^2 (\Delta Y/2) \; ,
\end{equation}

\begin{equation}
t_1 = - \frac{M^2}{2} \left[ 1 - \tanh (\Delta Y/2) \right] \; , 
\end{equation}

\begin{equation}
u_1 = - \frac{M^2}{2} \left[ 1 + \tanh (\Delta Y/2) \right] \; .
\end{equation}
Here, $ C_F = (N_c^2-1)/2 N_c $, $C_A = N_c$, $s$ is the squared center-of-mass
energy of the proton-proton system and $m$ is the heavy-quark mass; $\mu_{F_1}$,
$\mu_{F_2}$ are both set to $\sqrt{m_Q^2+\vec q^{\: 2}}$ and $\alpha_s$ is also
calculated at this scale. The upper and lower limits in the integration
over $\vec q^{\: 2}$ come from the constraint $M^2 \leq s$; in their
expression, $q^2_{\rm min}$ and $q^2_{\rm max}$ represent the kinematic cuts
on the heavy quark/antiquark transverse momentum. There is also a constraint on
$\Delta Y$, coming from the requirement that $M^2|_{\vec q^{\: 2}=0} \leq s$,
which is however always fulfilled for the values of $\Delta Y$ and $s$
considered in the numerical analysis presented below.

\section{Numerical analysis}
\label{numerical_analysis}

\subsection{Results}
\label{results}

In this Section we present our results for the dependence on the rapidity
interval between the two tagged bottom quarks, $\Delta Y \equiv y_1 - y_2$, of
the $\varphi$-averaged cross section $C_0$ and of the azimuthal correlations, $R_{n0}
= C_{n}/C_{0} \equiv \langle \cos n \varphi \rangle$, and their ratios, $R_{n0} = C_{n}/C_{m}$~\cite{Vera:2006un,Vera:2007kn}. We fix the $m_{1,2}$ masses at the value $m_b = 4.18$
GeV/$c^2$~\cite{Tanabashi:2018oca}. 

With the idea of matching realistic kinematic configurations, typical of the
current and possible future LHC analyses, we integrate the quark transverse
momenta in the symmetric range 20 GeV $< q_{1,2} <$ 100 GeV, fixing the
center-of-mass energy to $\sqrt{s} = 14$ TeV and studying the behavior of our
observables in the rapidity range $1.5 < \Delta Y < 9$. Ranges of the transverse
momenta of the bottom-jets ($b$-jets) are typical of CMS
analyses~\cite{Chatrchyan:2012dk,Chatrchyan:2012jua}.

\begin{figure}[p]
\centering
\includegraphics[scale=0.549,clip]{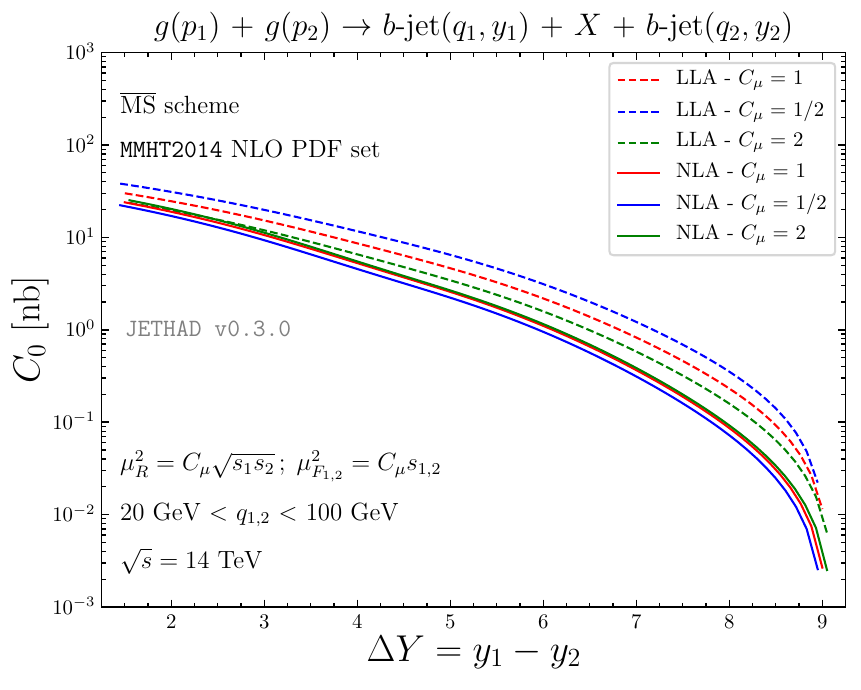}
\includegraphics[scale=0.55,clip]{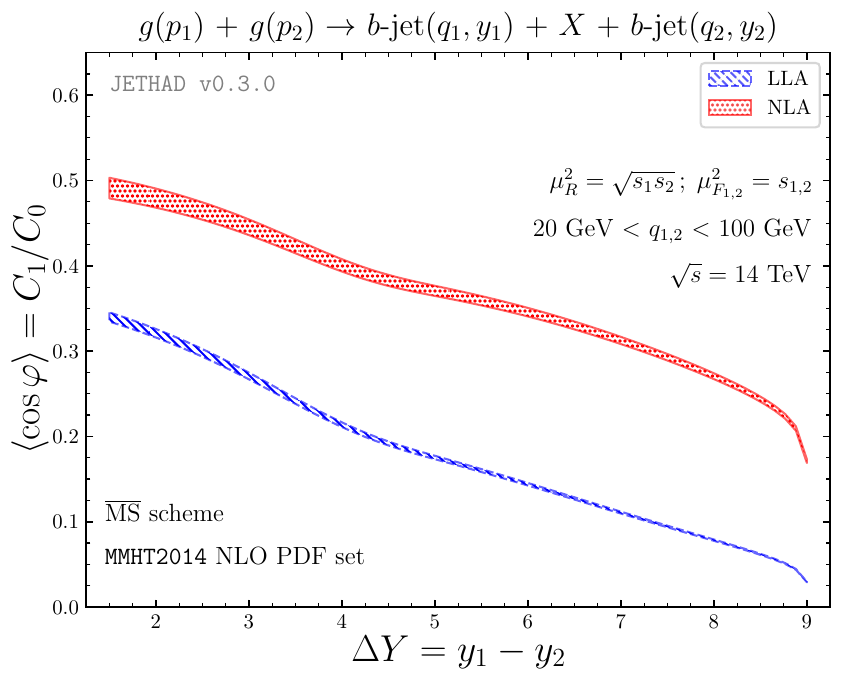}

\includegraphics[scale=0.55,clip]{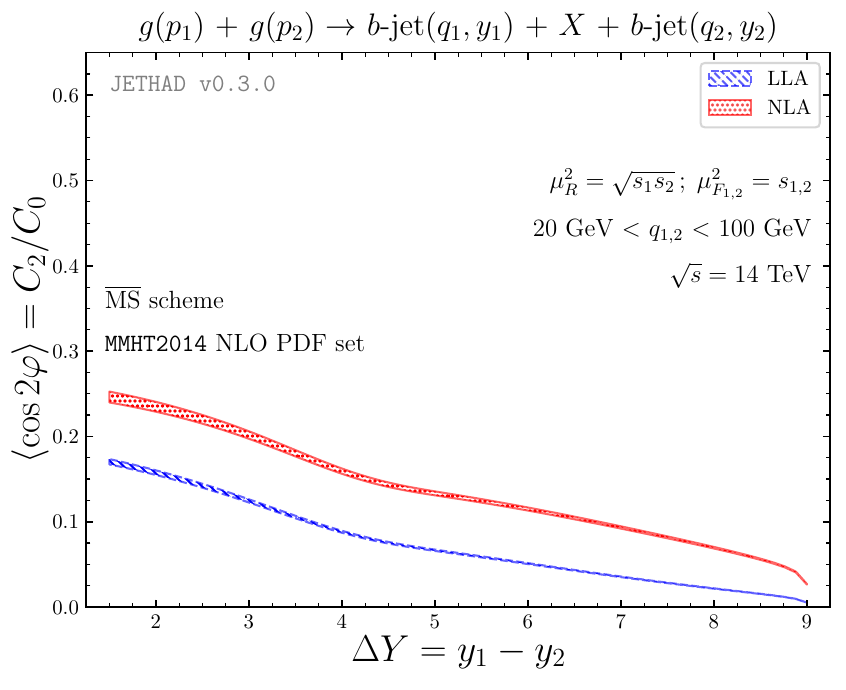}
\includegraphics[scale=0.55,clip]{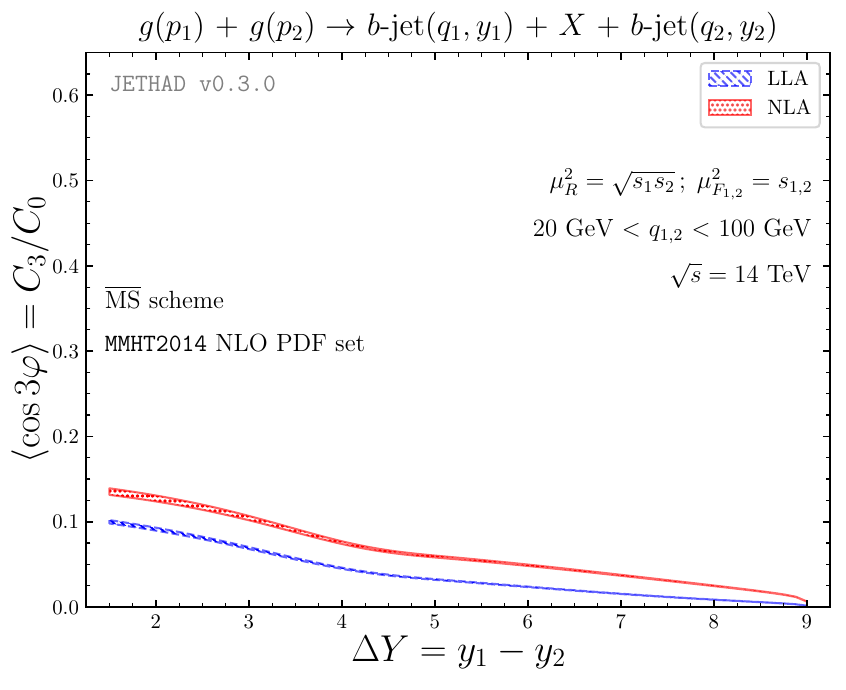}

\includegraphics[scale=0.55,clip]{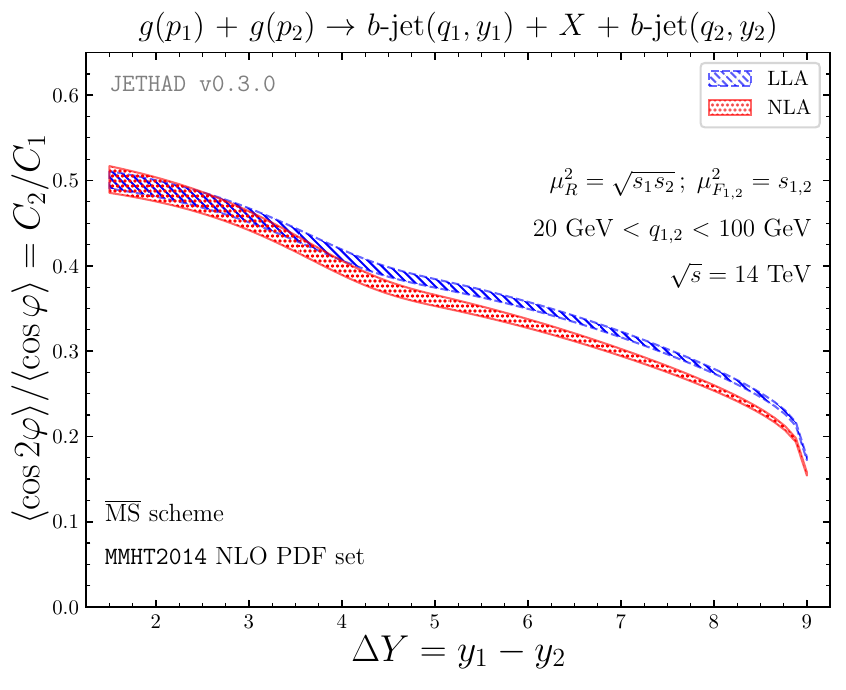}
\includegraphics[scale=0.55,clip]{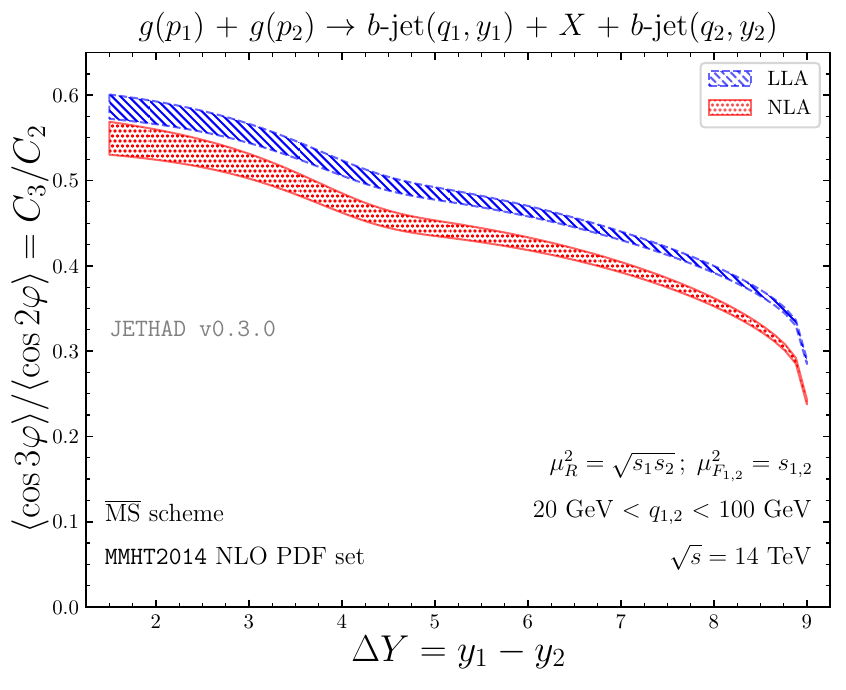}
\caption{$\Delta Y$-dependence of $C_0$ ($b$-jet pair) for different values of $C_\mu = \mu_R^2/\sqrt{s_1s_2}\equiv \mu^2_{F_{1,2}}/s_{1,2}$ (data points have been slightly shifted along the horizontal axis for the sake of readability), with $s_{1,2} = m^2_{1,2} + \vec{q}_{1,2}^{\; 2}$ and of several ratios $R_{nm} \equiv C_{n}/C_{m}$, for 20 GeV $< q_{1,2} <$ 100 GeV and $\sqrt{s} = 14$ TeV.}
\label{fig:C0-Rnm}
\end{figure}

Pure LLA and NLA BFKL predictions for the $\varphi$-averaged cross section,
$C_0$, together with the leading-order $p p \rightarrow q \bar{q}$ cross
section, are presented in Table~\ref{tab:C0-scales}.
Results for $C_0$ and for several azimuthal-correlation ratios, $R_{nm}$, are
shown in Fig.~\ref{fig:C0-Rnm}.

\begin{figure}[t]
\centering
\includegraphics[scale=0.549,clip]{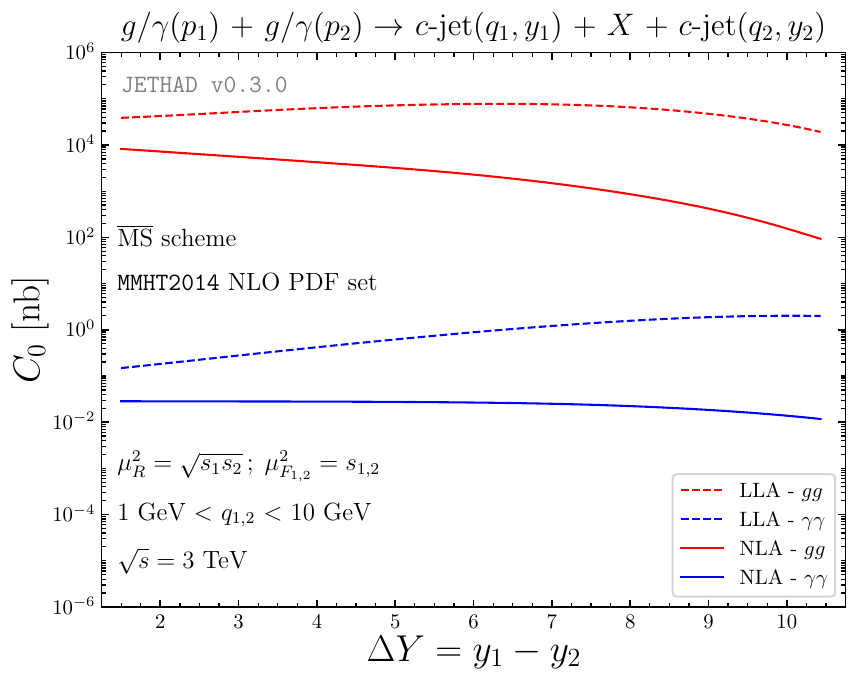}
\caption{$\Delta Y$-dependence of $C_0$ ($c$-jet pair) for both the hadroproduction ($gg$) and the photoproduction ($\gamma \gamma$) mechanisms, for 1 GeV $< q_{1,2} <$ 10 GeV and $\sqrt{s} = 3$ TeV.}
\label{fig:C0_hvg}
\end{figure}

As a complementary study (see Fig.~\ref{fig:C0_hvg}), we present results for
$C_0$ in the case of charmed-jet ($c$-jet) pair emission ($m_{1,2} = m_c = 1.2$
GeV/$c^2$), comparing predictions for the hadroproduction with the ones
related to the photoproduction mechanism (see Ref.~\cite{Celiberto:2017nyx}
for details on the theoretical framework and for a recent phenomenological
analysis conducted by some of us) in the kinematic configurations typical of
the future CLIC linear accelerator, namely 1 GeV $< q_{1,2} <$ 10 GeV, $\sqrt{s}
= 3$ TeV and $1.5 < \Delta Y < 10.5$.

All calculations are done in the $\overline{\rm MS}$ scheme.

\begin{table}[b]
\centering
\caption{$\Delta Y$-dependence of the $\varphi$-averaged cross section
  $C_0$ [nb] for $\sqrt{s} = 14$ TeV.
  $C_\mu$ stands for $\mu_R^2/\sqrt{s_1 s_2} \equiv \mu_{F_{1,2}}^2/s_{1,2}$.}
\label{tab:C0-scales}
\scriptsize
\begin{tabular}{r|lllllll}
\toprule
$\Delta Y$ &
$\tarr c \rm Box \\ Q \bar{Q} \earr$ &
$\tarr c \rm LLA \\ C_\mu = 1/2 \earr$ & 
$\tarr c \rm LLA \\ C_\mu = 1 \earr$ &
$\tarr c \rm LLA \\ C_\mu = 2 \earr$ &
$\tarr c \rm NLA \\ C_\mu = 1/2 \earr$ &
$\tarr c \rm NLA \\ C_\mu = 1 \earr$ & 
$\tarr c \rm NLA \\ C_\mu = 2 \earr$ \\
\midrule
1.5 & 33830.3 &  38.17(24) & 30.01(21) & 23.58(16) & 22.25(26) & 23.93(23) & 25.19(27) \\
3.0 & 3368.86 &  18.118(98) & 13.191(71) & 9.838(61) & 7.245(74) & 8.205(76) & 8.172(82) \\
4.5 & 124.333 &  6.996(33) & 4.715(23) & 3.276(16) &  2.209(20) & 2.411(17) & 2.422(19) \\       
6.0 & 3.19206 &  1.976(10) & 1.2430(60) & 0.8044(38) & 0.4497(35) & 0.4968(35) & 0.4868(37) \\
7.5 & 0.0610921 &  0.3317(16) & 0.19115(92) & 0.11509(57) & 0.05318(36) & 0.05785(39) & 0.05577(42) \\
9.0 & 0.000681608 & 0.02215(10) & 0.011458(56) & 0.006340(30) & 0.002566(17) & 0.002668(16) & 0.002513(16) \\
\bottomrule
\end{tabular}
\end{table}

\subsection{Numerical strategy and uncertainty estimate}
\label{numerical_strategy}

The numerical analysis was done using \textsc{Jethad}, a \textsc{Fortran} code
we recently developed, oriented towards the study of inclusive semi-hard
processes. In order to perform numerical integrations, \textsc{Jethad} was
interfaced with the \textsc{Cern} program library~\cite{cernlib} and with the
\textsc{Cuba} integrators~\cite{Cuba:2005,ConcCuba:2015}, making extensive use
of the \textsc{Vegas}~\cite{VegasLepage:1978} and the
\textsc{WGauss}~\cite{cernlib} integrators. The numerical stability of our
predictions was crosschecked using an independent \textsc{Mathematica} code.
The gluon PDFs ($f_{g_{1,2}}$) were calculated via the \textsc{MMHT2014} NLO PDF
parameterization~\cite{Harland-Lang:2014zoa} as implemented in the Les Houches
Accord PDF Interface (LHAPDF) 6.2.1~\cite{Buckley:2014ana}, while a two-loop
running coupling setup with $\alpha_s\left(M_Z\right)=0.11707$ with
dynamic-flavor thresholds was chosen.

The most relevant source of uncertainty, coming from the numerical
six-dimensional integration over the variables $|\vec q_1|$, $|\vec q_2|$,
$y_1$, $\nu$, $x_1$, and $x_2$, was directly estimated by \textsc{Vegas}.
Other sources of uncertainties, related with the upper cutoff in the $\nu$-
and the $\Delta$-integration in Eq.~(\ref{Cn}) and Eq.~(\ref{I4}),
respectively, are negligible with respect to the first one. Thus, the error
estimates of our predictions are just those given by \textsc{Vegas}. In order
to quantify the uncertainty related to the renormalization scale ($\mu_R$) and
the factorization one ($\mu_{F_{1,2}}$), we simultaneously vary the square of
both of them around their ``natural'' values, $\sqrt{s_1 s_2}$ and $s_{1,2}$
respectively, in the range 1/2 to two. The parameter $C_{\mu}$ entering
Table~\ref{tab:C0-scales} gives the ratio $C_{\mu} = \mu_R^2/\sqrt{s_1 s_2}
\equiv \mu_{F_{1,2}}^2/s_{1,2}$.

\subsection{Discussion}
\label{discussion}

The inspection of results for the $\varphi$-averaged cross section, $C_0$, in
the $b$-jet pair production case (Table~\ref{tab:C0-scales} and in the left
upper panel of Fig.~\ref{fig:C0-Rnm}) clearly indicates the usual onset of the
BFKL dynamics. On the one hand, although the high-energy resummation predicts
a growth with energy of the partonic-subprocess cross section, its convolution
with parent-gluon PDFs (Eq.~(\ref{dsigma_pp_conv})) leads, as a net effect, to
a falloff with $\Delta Y$ of both LLA and NLA predictions. On the other hand,
next-to-leading corrections to the BFKL kernel become more and more negative
when the rapidity distance grows, thus making NLA results steadily lower than
pure LLA ones.

Data in Table~\ref{tab:C0-scales} also show that the cross section $C_0$ is
smaller than the reference ``box'' cross section for small $\Delta Y$; at
larger rapidity differences, however, the BFKL mechanism with the gluonic
exchange in the $t$-channel starts to dominate. We stress, however, that for
our two heavy-quark (or two heavy-antiquark) tagged process, the ``box''
mechanism is not a background.

Azimuthal correlations (remaining panels of Fig.~\ref{fig:C0-Rnm}) are
always smaller than one and decrease when $\Delta Y$ grows (LLA results are
always more decorrelated than NLA ones), as an expected consequence of the
larger emission of undetected partons (the $X$ system in
Eq.~(\ref{numerical_analysis})). The cause for this narrowness, with respect to
other, recently investigated reactions, such as Mueller--Navelet jet, dihadron
or hadron-jet correlations (see, {\it e.g.},
Refs.~\cite{Caporale:2012ih,Celiberto:2017ptm,Bolognino:2018oth}), is
straightforward.
Since the two detected (anti)quarks stem from distinct vertices (each of them
together the respective antiparticle), their transverse momenta are
kinematically not constrained at all, even at leading order.

The analysis presented in Fig.~\ref{fig:C0_hvg} for the $c$-jet pair production
unambiguously shows that, at fixed center-of-mass energy and
transverse-momentum range, predictions for $C_0$ in the hadroproduction channel
($g g$) are several orders of magnitude higher than the corresponding ones in
the photoproduction case ($\gamma \gamma$). This comes as a result of two
competing effects. On one side, from a ``rough'' comparison between the
($g g$) impact factor (Eq.~(\ref{eq:imp.fac2})) and the ($\gamma \gamma$) one
(Eq.~(2) of Ref.~\cite{Celiberto:2017nyx} and footnote of
Ref.~\cite{Bolognino:2019ouc}), it emerges that the two analytic
structures are quite similar, the main difference being the fact that, since
the photon cannot interact directly with the Reggeized gluon, some terms
present in the first case are missing in the second one. In both of
the two impact factors there are constants that can be factorized out in the
final form of the cross section. Since two heavy-quark impact factors enter the
expression of cross sections, one has an overall factor
\begin{equation}
\label{k_had}
 \kappa_{(g g)} = \frac{\alpha_s^4 \, (N_c^2 - 1)}{(2 \pi N_c)^2}
\end{equation}
in the ($g g$) case and an overall factor
\begin{equation}
\label{k_gam}
 \kappa_{(\gamma \gamma)} = \frac{\alpha_{\rm em}^2 \, \alpha_s^2 \, e_c^4 (N_c^2 - 1)}{\pi^2}
\end{equation}
in the ($\gamma \gamma$) case, with $\alpha_{\rm em}$ the QED coupling and $e_c$
the electric charge of the charm quark in units of the positron charge. The
ratio between the two factors is
\begin{equation}
\label{k_had-gam}
 \kappa_{\rfrac{(g g)}{(\gamma \gamma)}} \equiv \frac{\kappa_{(g g)}}{\kappa_{(\gamma \gamma)}} \simeq 3 \div 4 \times 10^3 \;,
\end{equation}
which would explain the enhancement of the hadroproduction with respect to the
photoproduction.
On the other side, however, one should take into account the effect of the
parent-particle distributions: gluon PDF~\cite{Harland-Lang:2014zoa} or EPA
photon flux (see Eq.~(8) of Ref.~\cite{Celiberto:2017nyx}). It is possible to
show that the gluon PDF dominates over the photon flux in the moderate-$x$
region, while the second one prevails in the $x \to 0^+$ and $x \to 1^-$ limits.
In the realistic kinematic ranges we have considered in this paper and in
Refs.~\cite{Celiberto:2017nyx,Bolognino:2019ouc} the relevant $x$-region
turns to be just the intermediate one, thus leading to an enhancement
of the hadroproduction mechanism with respect to the photoproduction one
(see, {\it e.g.}, Fig.~\ref{fig:C0_hvg}) even larger than what suggested by
the ratio $\kappa_{\rfrac{(g g)}{(\gamma \gamma)}}$.

\section{Summary and outlook}
\label{summary_outlook}

We have proposed the inclusive hadroproduction of two heavy quarks separated by
a large rapidity interval as a new channel for the investigation of BFKL
dynamics. We have performed an all-order resummation of the leading energy
logarithms and a resummation of the next-to-leading ones entering the BFKL
Green's function. In this approximation, the cross section can be written
as the convolution of the partonic cross section for the collision of two
gluons producing the two heavy quarks with the respective gluon PDFs.

We have calculated the cross section for this process summed over the relative
azimuthal angle of the two tagged quarks and presented results for the
azimuthal angle correlations. The behavior of these observables turned to be
the usual one, characteristic feature of the onset of the BFKL dynamics.
Finally, a comparison between the photoproduction and the hadroproduction
mechanism has been carried out. 

This process enriches the selection of semi-hard reactions that can be used as
probes of the QCD in the high-energy limit, and in particular of the BFKL
resummation mechanism, in the kinematic ranges of the LHC and of future
hadronic colliders. 

Several prospective developments of this work can be planned and afforded. 
The first one consists in the calculation of the NLO correction to the forward
heavy-quark pair impact factor, which would allow for a full NLA BFKL treatment
of the process under consideration. 
The second one is to include into the theoretical analysis the quark
fragmentation needed to match, from the theoretical side, the experimental
tagging procedure of heavy-quark mesons.
Since the photoproduction channel has already been considered
(Refs.~\cite{Celiberto:2017nyx,Bolognino:2019ouc}), a process of
photo/hadro-production (when the first (anti)quark is emitted by a (quasi-)real
photon, while the second one stems from a gluon), hybrid with respect to the
previous ones, can also be examined. 
One last idea is to investigate semi-hard channels featuring the emission of a
single quark. For instance, one can study the single forward heavy-quark
production, convolving the corresponding impact factor with the unintegrated
gluon density (UGD) in the proton.

\begin{figure}[t]
\centering
\includegraphics[width=1.0\textwidth]{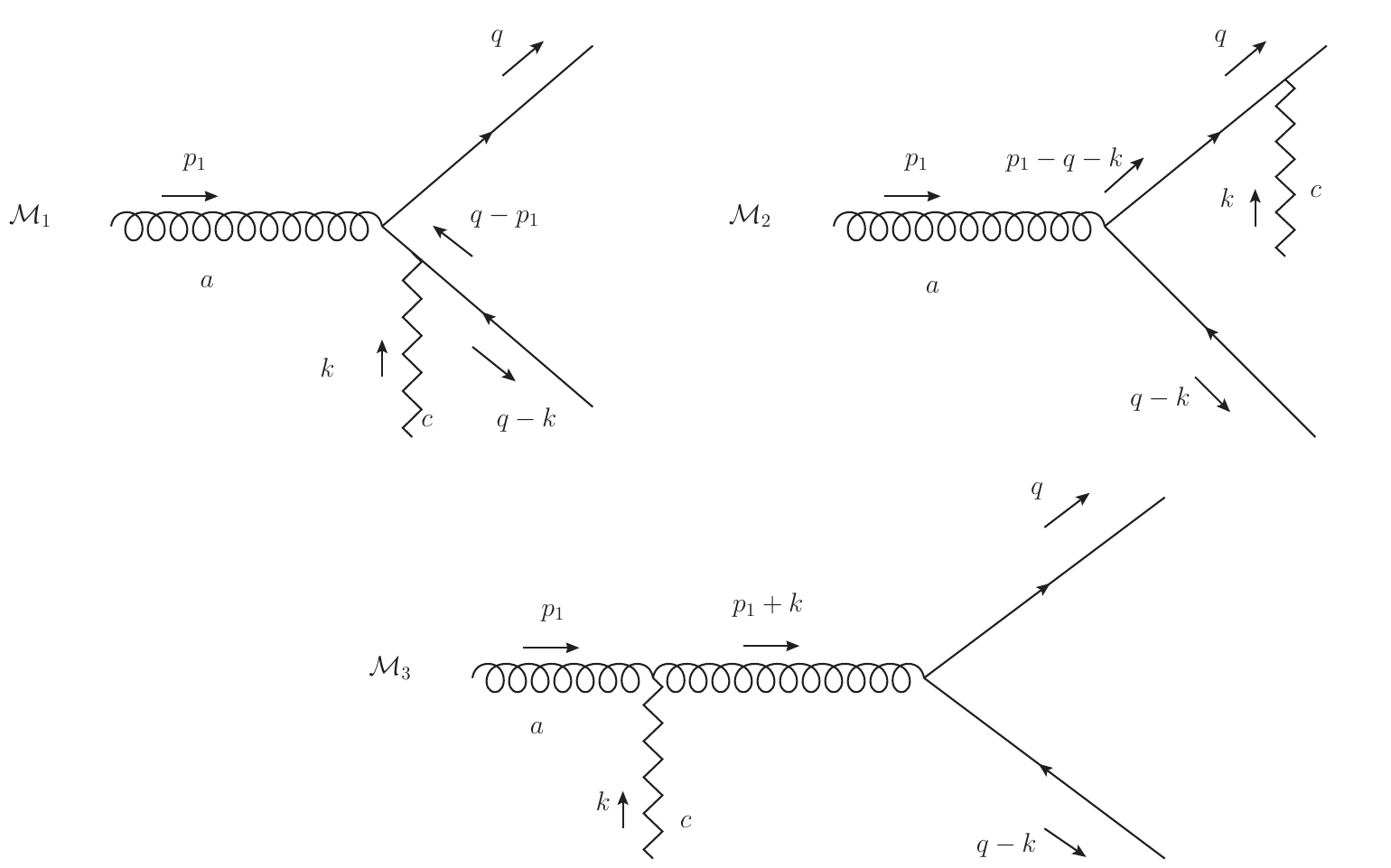}
\caption{Feynman diagrams relevant for the calculation of the impact factor for
the heavy-quark pair hadroproduction. The zigzag line denotes a Reggeized
gluon.}
\label{fig:hadroproduction_IF}
\end{figure}

\label{summary_outlook}

\hypertarget{app:IF_def-link}{
\section*{Appendix~A}}

In this Section we give the expression of the leading-order impact factor,
together with the functional form of the amplitude for the
$g + R \to q \bar{q}$ subprocess, where $R$ here means ``Reggeized gluon''.
The leading-order impact factor is defined as~\cite{Fadin:1998fv}
\begin{equation}
\begin{split}
  d\Phi_{gg}^{\lbrace{Q\bar{Q}\rbrace}}(\vec{q},\vec{k},z) = & \frac{\braket{cc'|\mathcal{\widehat{P}}|0}}{2\left(N^2-1\right)}
 \\ &  \times  \sum_{\lambda_Q\lambda_{\bar{Q}}\lambda_G}\sum_{Q\bar{Q}a} \int\frac{ds_{gR}}{2\pi}d\rho_{\lbrace{Q\bar{Q}\rbrace}}\Gamma_{g \rightarrow \lbrace{Q\bar{Q}\rbrace}}^{ca}\left(q,k,z\right)\left(\Gamma_{g \rightarrow \lbrace{Q\bar{Q}\rbrace}}^{ac'}\left(q,k,z\right)\right)^{\ast} \;,
\end{split}
\label{eq:imp.fac}
\end{equation}
where 
\begin{equation}
\braket{cc'|\mathcal{\widehat{P}}|0} = \frac{\delta^{cc'}}{\sqrt{N^2-1}}
\end{equation}
is the projector on the singlet state. We take the sum over helicities,
$\{\lambda_Q,\lambda_{\bar{Q}}\}$, and over color indices, $\{Q,\bar{Q}\}$, of
the two produced particles (quark and antiquark) and average over polarization
and color states of the incoming gluon. In Eq.~(\ref{eq:imp.fac}), $s_{gR}$
denotes the invariant squared mass of the gluon-Reggeon system, while $
d\rho_{\lbrace{Q\bar{Q}\rbrace}}$ is the differential phase space of the outgoing
particles. The amplitude $\Gamma_{g \rightarrow \lbrace{Q\bar{Q}\rbrace}}^{ca}$ describes
the production of quark-antiquark pair in a collision between a gluon and a
Reggeon. The latter can be treated as an ordinary gluon in the so called
``nonsense'' polarization state $\epsilon_R^\mu= p_2^\mu/s$, see, {\it e.g.},
Ref.~\cite{Fadin:2001dc}.
Having two particles produced in the intermediate state, one can write
\begin{equation}
\label{phasespace}
\frac{ds_{gR}}{2 \pi} d\rho_{\lbrace{Q\bar{Q}\rbrace}} = \frac{1}{2\left(2 \pi \right)^3} \delta \left(1-z-\bar{z}\right) \delta^{(2)} \left( \vec{k}-\vec{q}-\vec{\bar{q}} \right) \frac{dz d\bar{z}}{z \bar{z}} d^2\vec{q} \; d^2\vec{\bar{q}} \; , 
\end{equation}
with $\bar{q}$ the antiquark momentum, and $\bar{z}$ its longitudinal
momentum fraction (with respect to the incoming gluon).
Summing over the three contributions, $\{{\cal M}_{1,2,3}\}$, of
Fig.~\ref{fig:hadroproduction_IF}, one gets
\begin{equation}
\label{amp}
\begin{split}
\Gamma_{g \rightarrow \lbrace{Q\bar{Q}\rbrace}}^{ca} & = ig^2\left(\tau^{a}\tau^{c}\right)\bar{u}\left(q\right)\left(mR\slashed{\epsilon} - 2 z \vec{P} \cdot \vec{\epsilon} - \vec{\slashed{P}} \slashed{\epsilon}\right)\frac{\slashed{p}_2}{\hat{s}}v\left(\bar{q}\right) \\ & + ig^2\left(\tau^{c}\tau^{a}\right)\bar{u}\left(q\right)\left(m\bar{R}\slashed{\epsilon} - 2 z \vec{\bar{P}} \cdot \vec{\epsilon}- \vec{\slashed{\bar{P}}} \; \; \slashed{\epsilon} \right)\frac{\slashed{p}_2}{\hat{s}}v\left(\bar{q} \right) \; ,
\end{split}
\end{equation}
where $\hat{s} = W^2$, $\epsilon^\mu$ identifies the gluon polarization vector, $\{\tau\}$ are
the $SU(3)$ color matrices and $R, \bar{R}, \vec{P}, \vec{\bar{P}}$ are
defined in Eqs.~(\ref{laR})-(\ref{LaPbar}), respectively.
Using Eq.~(\ref{amp}) together with Eq.~(\ref{eq:imp.fac}) and performing
sums and integrations (the latter ones only on the antiquark variables), our
final result reads
\begin{equation}
\begin{split}
d\Phi^{\lbrace{Q\bar{Q}\rbrace}}_{gg}(\vec{k},\vec{q},z)& =\frac{\alpha_s^2
\sqrt{N_c^2-1}}{2\pi N_c}\left[\left(m^2\left(R+\bar{R}\right)^2
+\left(z^2+\bar{z}^2 \right)\left(\vec{P}+\vec{\bar{P}}\right)^2\right)
\right. \\ & \left. -\frac{N_c^2}{N_c^2-1}\left(2m^2R\bar{R}
+\left(z^2+\bar{z}^2\right)2\vec{P} \cdot \vec{\bar{P}}\right)\right]
\; d^2\vec{q} \; dz\;,
\end{split}
\end{equation}
which exactly matches the definition of the impact factor given in
Eq.~(\ref{eq:imp.fac2}).

\hypertarget{app:IF_calc-link}{
\section*{Appendix~B}}
\label{appendix}

In this Section we present the four integrals necessary to perform the
$(n, \nu)$-projection of the impact factor.  

The first integral,
\begin{equation}
I_1\equiv\int\frac{d^2\vec{k}}{\pi\sqrt{2}}\left(\vec{k}^{\;2}\right)^{i\nu-\frac{3}{2}}e^{in\theta} \; , \quad {\rm{for}} \; n \neq 0 \;,    
\end{equation}
vanishes because of the periodicity condition on the angle $\theta$.

The second integral reads
\begin{equation}
\begin{split}
I_2\left(\lambda\right) \equiv \int\frac{d^2\vec{k}}{\pi\sqrt{2}}(\vec{k}^{\;2})^{i\nu-\frac{3}{2}}e^{in\theta}\frac{(\vec{k}^{\;2})^{\lambda}}{m^2+(\vec{q}-\vec{k})^2}
\end{split}
\end{equation}
\begin{equation}
\begin{split}
& = \frac{\left(\vec{q}^{\; 2}\right)^{\frac{n}{2}}e^{in\varphi}}{\sqrt{2}}\frac{1}{\left(m^2+\vec{q}^{\;2}\right)^{\frac{3}{2}+\frac{n}{2}-i\nu-\lambda}} \frac{\Gamma\left(\frac{1}{2}+\frac{n}{2}+i\nu+\lambda\right)\Gamma\left(\frac{1}{2}+\frac{n}{2}-i\nu-\lambda\right)}{\Gamma\left(1+n\right)} \\ & \times \frac{\left(\frac{1}{2}+\frac{n}{2}-i\nu-\lambda\right)}{\left(-\frac{1}{2}+\frac{n}{2}+i\nu+\lambda\right)}\; _2F_1\left(-\frac{1}{2}+\frac{n}{2}+i\nu+\lambda,\frac{3}{2}+\frac{n}{2}-i\nu-\lambda,1+n,\zeta\right) \;.
\end{split}
\end{equation}

The third integral can be presented as
\begin{equation}
\label{I3_start}
 I_3
 \equiv \int\frac{d^2\vec{k}}{\pi\sqrt2}(\vec{k}^{\,\,2})^{i\nu-\frac{3}{2}}e^{in\theta}
 \frac{1}{\left(m^2+(\vec{q}-\vec{k})^2\right)^2} 
\end{equation}
\begin{equation}
\begin{split}
= & \frac{\left(\vec{q}^{\; 2}\right)^{\frac{n}{2}}e^{in\varphi}}{\sqrt{2}} \frac{1}{\left(m^2+\vec{q}^{\;2}\right)^{\frac{5}{2}+\frac{n}{2}-i\nu}} \frac{\Gamma\left(\frac{1}{2}+\frac{n}{2}+i\nu\right)\Gamma\left(\frac{1}{2}+\frac{n}{2}-i\nu\right)}{\Gamma\left(1+n\right)}\frac{\left(\frac{1}{2}+\frac{n}{2}-i\nu\right)}{\left(-\frac{1}{2}+\frac{n}{2}+i\nu\right)}\; \\ & \times \left(\frac{3}{2}+\frac{n}{2}-i\nu\right)\; _2F_1\left(-\frac{1}{2}+\frac{n}{2}+i\nu,\frac{5}{2}+\frac{n}{2}-i\nu,1+n,\zeta \right) \;.
\end{split} 
\end{equation}
For the sake of completeness, we show the entire derivation of the fourth integral, defined as
\begin{equation}
I_4\left(\lambda\right) \equiv \int\frac{d^2\vec{k}}{\pi\sqrt{2}} (\vec{k}^{\;2})^{i\nu-\frac{3}{2}}e^{in\theta}\frac{(\vec{k}^{\;2})^{\lambda}}{(m^2+(\vec{q}-\vec{k})^2)(m^2+(\vec{q}-z \vec{k})^2)} \;,
\end{equation}
which is the most cumbersome one. The strategy to calculate $I_2(\lambda)$ and $I_3$ is the same. To lighten the notation, it is useful to define $\alpha=i\nu+\lambda$, then the integral can be easily put in the following form:
\begin{equation}
I_4\left(\lambda\right)=\frac{1}{z^2} \int\frac{d^2\vec{k}}{\pi\sqrt{2}}\frac{(\vec{k}\cdot\vec{l}\;)^n}{(\vec{k}^{\;2})^{\frac{3}{2}+\frac{n}{2}-\alpha}\left(m^2+\left(\vec{q}-\vec{k}\right)^2\right)\left(\frac{m^2}{z^2}+\left(\frac{\vec{q}}{z}-\vec{k}\right)^2\right)} \;.
\end{equation}
where $\vec{l} \equiv (i,1)$ and the following formula,
\begin{equation}
e^{in\theta}=\left(\cos{\theta}+i\sin{\theta}\right)^n=\left(\frac{k_x+ik_y}{|\vec{k}\;|}\right)^n=\frac{(\vec{k}\cdot\vec{l}\;)^n}{(\vec{k}^{\;2})^{\frac{n}{2}}} \;,
\end{equation}
has been used. Using the Feynman parametrization
\begin{equation}
\label{feynman param.2}
\begin{split}
\frac{1}{s^Mt^Nw^L}&=\frac{\Gamma\left(M+N+L\right)}{\Gamma\left(M\right)\Gamma\left(N\right)\Gamma\left(L\right)}\int_0^1 dx \int_0^1 dy \int_0^1 dz \frac{x^{M-1}y^{N-1}z^{L-1}\delta\left(1-x-y-z\right)}{\left(xs+yt+zw\right)^{M+N+L}} \\ & =\frac{\Gamma\left(M+N+L\right)}{\Gamma\left(M\right)\Gamma\left(N\right)\Gamma\left(L\right)}\int_0^1 dx \int_0^{1-x} dy \frac{x^{M-1}y^{N-1}\left(1-x-y\right)^{L-1}}{\left(xs+yt+\left(1-x-y\right)w\right)^{M+N+L}} \;.
\end{split}
\end{equation}
one finds that
\begin{equation}
\begin{split}
& I_4\left(\lambda\right)= \frac{1}{z^2}\frac{\Gamma\left(\frac{7}{2}-\alpha+\frac{n}{2}\right)}{\Gamma\left(\frac{3}{2}-\alpha+\frac{n}{2}\right)} \int \frac{d^2\vec{k}}{\pi\sqrt{2}} \\& \times \int_0^1 dx \int_0^{1-x} dy \frac{\left(1-x-y\right)^{\frac{1}{2}-\alpha+\frac{n}{2}}\left(\vec{k} \cdot \vec{l} \; \right)^n}{\left[x\left(\frac{m^2}{z^2}+\left(\frac{\vec{q}}{z}-\vec{k}\right)^2\right)+y\left(m^2+\left(\vec{q}-\vec{k}\right)^2\right)+\left(1-x-y\right)\vec{k}^{\; 2}\right]^{\frac{7}{2}-\alpha+\frac{n}{2}}} \;.
\end{split}
\end{equation}
Making the following substitution, $\vec{k} \to \vec{k}+\left(\frac{x}{z}+y\right)\vec{q}$, and observing that 
\begin{equation}
\left(\vec{k}\cdot \vec{l}+\left(\frac{x}{z}+y\right) \vec{q}\cdot \vec{l}\;\right)^n = \sum_j{n\choose j}\left(\vec{k} \cdot \vec{l}\;\right)^j \left(\frac{x}{z}+y\right)^{n-j}\left(\vec{q} \cdot \vec{l}\;\right)^{n-j} \;,
\end{equation}
after some trivial calculations, one obtains
\begin{equation}
\label{eq:binomial2}
\begin{split}
I_4\left(\lambda\right)=& \frac{1}{z^2}\left(\frac{5}{2}-\alpha+\frac{n}{2}\right)\left(\frac{3}{2}-\alpha+\frac{n}{2}\right) \int_0^1 dx \int_0^{1-x} dy \left(1-x-y\right)^{\frac{1}{2}-\alpha+\frac{n}{2}} \\ & \times \int \frac{d^2\vec{k}}{\sqrt{2}}\frac{\sum_j{n\choose j}\left(\vec{k} \cdot \vec{l}\;\right)^j \left(\frac{x}{z}+y\right)^{n-j}\left(\vec{q} \cdot \vec{l}\;\right)^{n-j}}{\left[\vec{k}^{\; 2}+L^2 \right]^{\frac{7}{2}-\alpha+\frac{n}{2}}} \;,
\end{split}
\end{equation}
where 
\begin{equation}
L^2=\left(\frac{x}{z^2}+y\right)\left(m^2+\vec{q}^{\; 2}\right)-\left(\frac{x}{z}+y\right)^2\vec{q}^{\; 2} \;.
\end{equation}
The only term which gives non-zero contribute in the binomial in
Eq.~(\ref{eq:binomial2}) is the $0^{\rm th}$ coefficient, namely
\begin{equation}
\left(\frac{x}{z}+y\right)^n\left(\vec{q}\cdot\vec{l}\;\right)^n=\left(\frac{x}{z}+y\right)^n\left(\vec{q}^{\;2}\right)^{\frac{n}{2}}e^{in\varphi} \;,
\end{equation}
where $\varphi$ is the azimuthal angle of the vector $\vec{q}$.

Hence,
\begin{equation}
\begin{split}
I_4 & \left(\lambda\right)= \frac{\left(\vec{q}^{\; 2}\right)^{\frac{n}{2}}e^{in\varphi}}{z^2}\left(\frac{5}{2}-\alpha+\frac{n}{2}\right)\left(\frac{3}{2}-\alpha+\frac{n}{2}\right) \\ & \times \int_0^1 dx \int_0^{1-x} dy \left(1-x-y\right)^{\frac{1}{2}-\alpha+\frac{n}{2}}\left(\frac{x}{z}+y\right)^n \int \frac{d^2\vec{k}}{\sqrt{2}}\frac{1}{\left[\vec{k}^{\; 2}+L^2 \right]^{\frac{7}{2}-\alpha+\frac{n}{2}}} \;.
\end{split}
\end{equation}
For the integration in $d^2\vec{k}$, we use the formula 
\begin{equation}
\label{integral_k}
\int \frac{d^{2}k}{(2\pi)^{3}}\frac{1}{(\vec{k}^{\,\,2} + L^2)^\rho}=\frac{2}{(4\pi)^{2}} \frac{\Gamma(\rho-1)}{\Gamma(\rho)}(L^2)^{-\rho+1} \; ,
\end{equation}
setting $\rho = \frac{7}{2}-\alpha+\frac{n}{2}$. We then obtain
\begin{equation}
\begin{split}
I_4\left(\lambda\right)=& \frac{\left(\vec{q}^{\; 2}\right)^{\frac{n}{2}}e^{in\varphi}}{z^2\sqrt{2}}\frac{\left(\frac{3}{2}-\alpha+\frac{n}{2}\right)}{\left(m^2+\vec{q}^{\; 2}\right)^{\frac{5}{2}-\alpha+\frac{n}{2}}} \int_0^1 dx \int_0^{1-x} dy \\ & \times \left(1-x-y\right)^{\frac{1}{2}-\alpha+\frac{n}{2}}\left(\frac{x}{z}+y\right)^n \left[\left(\frac{x}{z^2}+y\right)-\zeta\left(\frac{x}{z}+y\right)^2\right]^{-\frac{5}{2}+\alpha-\frac{n}{2}} \;,
\end{split}
\end{equation}
where
\begin{equation}
\label{zeta}
\zeta=\frac{\vec{q}^{\;2}}{m^2+\vec{q}^{\;2}} \;.
\end{equation}
To integrate this expression over one of the two Feynman parameters, we perform
the following change of variables:
\begin{align}
\label{transf.}
 & x \,=\, \tau \Delta \; ,
 \\ \nonumber 
 & y \,=\, \tau \left(1-\Delta \right) \;,
\end{align}  
The Jacobian determinant is simply 
\begin{equation}
||J||=\tau \; ,
\end{equation}
and the integral $I_4\left(\lambda\right)$ becomes
\begin{equation}
\begin{split}
I_4\left(\lambda\right)= \frac{\left(\vec{q}^{\; 2}\right)^{\frac{n}{2}}e^{in\varphi}}{z^2\sqrt{2}} & \frac{\left(\frac{3}{2}-\alpha+\frac{n}{2}\right)}{\left(m^2+\vec{q}^{\; 2}\right)^{\frac{5}{2}-\alpha+\frac{n}{2}}} \int_0^1 d\Delta \left(1+\frac{\Delta}{z}-\Delta\right)^n \left(1+\frac{\Delta}{z^2}-\Delta\right)^{-\frac{5}{2}+\alpha-\frac{n}{2}} \\ & \times \int_0^1 d\tau \left(1-\tau\right)^{\frac{1}{2}-\alpha+\frac{n}{2}} \tau^{-\frac{3}{2}+\alpha+\frac{n}{2}} \left(1-A\tau\right)^{-\frac{5}{2}+\alpha-\frac{n}{2}} \;,
\end{split}
\end{equation}
where
\begin{equation}
A=\zeta \frac{\left(1+\frac{\Delta}{z}-\Delta\right)^2}{\left(1+\frac{\Delta}{z^2}-\Delta\right)} \; .
\end{equation}
Let us now consider the integral representation of the hypergeometric function, 
\begin{equation}
\label{hypergeometric}
\mathcal{B}(b,c-b) \, {_2}F_1(a,b,c,z) = \int_0^1 dx \, x^{b-1} (1-x)^{c-b-1} (1-zx)^{-a} \;,
\end{equation}
with the Euler function 
\begin{equation}
\label{beta_function}
\mathcal{B}(u,\omega) = 
\frac{\Gamma(u)\Gamma(\omega)}{\Gamma(u+\omega)}
\; . 
\end{equation}
For the $\tau$ integration it is enough to consider Eq.~(\ref{hypergeometric})
and the Eq.~(\ref{beta_function}), setting:
\begin{equation*}
a \,=\, \frac{5}{2}-\alpha+\frac{n}{2} \; , \quad \quad b \,=\, -\frac{1}{2}+\alpha+\frac{n}{2} \; , \quad \quad c \,=\, 1+n \; , \quad \quad  z \,=\, A \; .
\end{equation*}
Finally, expressing $A$ and $\alpha$ in their explicit form and making use of
the property
\[
_2F_1(a,b,c,z)= \; _2F_1(b,a,c,z) \;,
\]
one finds
\begin{equation}
\begin{split}
I_4\left(\lambda\right) = & \frac{\left(\vec{q}^{\; 2}\right)^{\frac{n}{2}}e^{in\varphi}}{z^2\sqrt{2}}  \frac{\left(\frac{3}{2}-i\nu-\lambda+\frac{n}{2}\right)}{\left(m^2+\vec{q}^{\; 2}\right)^{\frac{5}{2}-i\nu-\lambda+\frac{n}{2}}} \frac{\Gamma\left(\frac{1}{2}+\frac{n}{2}+i\nu+\lambda \right)\Gamma\left(\frac{1}{2}+\frac{n}{2}-i\nu-\lambda \right)}{\Gamma\left(1+n\right)} \\ & \times \frac{\left(\frac{1}{2}+\frac{n}{2}-i\nu-\lambda \right)}{\left(-\frac{1}{2}+\frac{n}{2}+i\nu+\lambda \right)} \int_0^1 d\Delta \left(1+\frac{\Delta}{z}-\Delta\right)^n \left(1+\frac{\Delta}{z^2}-\Delta\right)^{-\frac{5}{2}+i\nu+\lambda-\frac{n}{2}} \; \\ & \times \; _2F_1\left(-\frac{1}{2}+i\nu+\lambda+\frac{n}{2},\frac{5}{2}-i\nu-\lambda+\frac{n}{2},1+n,\zeta \frac{\left(1+\frac{\Delta}{z}-\Delta\right)^2}{\left(1+\frac{\Delta}{z^2}-\Delta\right)}\right) \;.
\end{split}
\end{equation}

\section*{Acknowledgments}

We thank G.~Krintiras and C.~Royon for fruitful discussions.
\\
F.G.C. acknowledges support from the Italian Ministry of Education,
University and Research under the FARE grant ``3DGLUE'' (n. R16XKPHL3N). 
\\
D.I. thanks the Dipartimento di Fisica dell'U\-ni\-ver\-si\-t\`a della Calabria
and the Istituto Nazio\-na\-le di Fisica Nucleare (INFN), Gruppo collegato di
Cosenza, for the warm hospitality and the financial support.


\begin{thebibliography}{99}

\bibitem{Gribov:1984tu}
  L.V.~Gribov, E.M.~Levin, M.G.~Ryskin,
  Phys.\ Rept.\  {\bf 100} (1983) 1.

\bibitem{BFKL}
V.S.~Fadin, E.~Kuraev, L.~Lipatov, Phys. Lett. B \textbf{60}, 50 (1975);
%
Sov. Phys. JETP \textbf{44}, 443 (1976);
%
E.~Kuraev, L.~Lipatov, V.S.~Fadin, Sov. Phys. JETP \textbf{45}, 199 (1977);
%
I.~Balitsky, L.~Lipatov, Sov. J. Nucl. Phys. \textbf{28}, 822 (1978).

\bibitem{Fadin:1998py}
V.S.~Fadin, L.N.~Lipatov,
Phys.\ Lett.\ B {\bf 429} (1998) 127.

\bibitem{Ciafaloni:1998gs}
M.~Ciafaloni, G.~Camici,
Phys.\ Lett.\ B {\bf 430} (1998) 349.

\bibitem{Fadin:1998jv}
V.S.~Fadin, R.~Fiore, A.~Papa, Phys.\ Rev.\ D {\bf 60} (1999) 074025.

\bibitem{FG00}
V.S. Fadin, D.A. Gorbachev, Pisma v Zh. Eksp. Teor. Fiz. {\bf 71} (2000) 322
[JETP Letters {\bf 71} (2000) 222]; Phys. Atom. Nucl. {\bf 63} (2000) 2157
[Yad. Fiz. {\bf 63} (2000) 2253].

\bibitem{FF05}
V.S.~Fadin, R.~Fiore, Phys. Lett. {\bf B610} (2005) 61
[{\it Erratum-ibid.} {\bf 621} (2005) 61];
Phys. Rev. D {\bf 72} (2005) 014018.

\bibitem{fading}
V.S.~Fadin, R.~Fiore, M.I.~Kotsky, A.~Papa, Phys. Lett. {\bf D61} (2000)
094005.

\bibitem{fadinq}
V.S.~Fadin, R.~Fiore, M.I.~Kotsky, A.~Papa, Phys. Lett. {\bf D61} (2000).

\bibitem{Cia}
M.~Ciafaloni, D.~Colferai, Nucl. Phys. B {\bf 538} (1999) 187.

\bibitem{Ciafaloni:2000sq}
M.~Ciafaloni, G.~Rodrigo, JHEP {\bf 0005} (2000) 042.

\bibitem{bar1}
J.~Bartels, D.~Colferai, G.P.~Vacca,
Eur.\ Phys.\ J.\  C {\bf 24} (2002) 83.

\bibitem{bar2}
J.~Bartels, D.~Colferai, G.P.~Vacca,
Eur.\ Phys.\ J.\  C {\bf 29} (2003) 235.

\bibitem{Caporale:2011cc}
F.~Caporale, D.Yu.~Ivanov, B.~Murdaca, A.~Papa, A.~Perri,
JHEP {\textbf 1202}, 101 (2012). 

\bibitem{Ivanov:2012ms}
D.Yu.~Ivanov, A.~Papa,
JHEP {\bf 1205}, 086 (2012).

\bibitem{Colferai:2015zfa}
D.~Colferai, A.~Niccoli, JHEP {\bf 1504}, 071 (2015).

\bibitem{Ivanov:2012iv}
D.Yu.~Ivanov, A.~Papa,
JHEP {\bf 1207} (2012) 045
[arXiv:1205.6068 [hep-ph]].

\bibitem{IKP04}
D.Yu. Ivanov, M.I.~Kotsky, A. Papa, Eur. Phys. J. C {\bf 38} (2004) 195.

\bibitem{gammaIF}
J.~Bartels, S.~Gieseke, C.F.~Qiao, Phys. Rev. D 63 (2001) 056014
[{\it Erratum-ibid.} D 65 (2002) 079902];
J.~Bartels, S.~Gieseke, A.~Kyrieleis, Phys. Rev. D 65 (2002) 014006;
J.~Bartels, D.~Colferai, S.~Gieseke, A.~Kyrieleis,
Phys. Rev. D 66 (2002) 094017;
J.~Bartels, Nucl. Phys. (Proc. Suppl.) (2003) 116;
J.~Bartels, A.~Kyrieleis, Phys. Rev. D 70 (2004) 114003;
V.S.~Fadin, D.Yu.~Ivanov, M.I.~Kotsky, Phys. Atom. Nucl. 65 (2002)
1513 [Yad. Fiz. 65 (2002) 1551]; Nucl. Phys. B 658 (2003) 156.

\bibitem{Balitsky2012}
I.~Balitsky, G.A.~Chirilli, Phys. Rev. D 87 (2013) 014013.

\bibitem{Celiberto:2017ius}
  F.G.~Celiberto, PhD thesis,
  arXiv:1707.04315 [hep-ph].

\bibitem{Bolognino:2018rhb}
  A.D.~Bolognino, F.G.~Celiberto, D.Yu.~Ivanov, A.~Papa,
  Eur.\ Phys.\ J.\ C {\bf 78} (2018) no.12,  1023
  [arXiv:1808.02395 [hep-ph]].

\bibitem{Bolognino:2018mlw}
  A.D.~Bolognino, F.G.~Celiberto, D.Yu.~Ivanov, A.~Papa,
  arXiv:1808.02958 [hep-ph].

\bibitem{Bolognino:2019bko}
  A.D.~Bolognino, F.G.~Celiberto, D.Yu.~Ivanov, A.~Papa,
  arXiv:1902.04520 [hep-ph].

\bibitem{Ivanov:2004pp}
  D.Yu.~Ivanov, M.I.~Kotsky, A.~Papa,
  Eur.\ Phys.\ J.\ C {\bf 38} (2004) 195
  [hep-ph/0405297].

\bibitem{Ivanov:2005gn}
  D.Yu.~Ivanov, A.~Papa,
  Nucl.\ Phys.\ B {\bf 732} (2006) 183
  [hep-ph/0508162].

\bibitem{Ivanov:2006gt}
  D.Yu.~Ivanov, A.~Papa,
  Eur.\ Phys.\ J.\ C {\bf 49} (2007) 947
  [hep-ph/0610042].

\bibitem{Enberg:2005eq}
  R.~Enberg, B.~Pire, L.~Szymanowski, S.~Wallon,
  Eur.\ Phys.\ J.\ C {\bf 45} (2006) 759
   Erratum: [Eur.\ Phys.\ J.\ C {\bf 51} (2007) 1015]
  [hep-ph/0508134].

\bibitem{Mueller:1986ey}
A.H.~Mueller, H.~Navelet, 
Nucl. Phys. B \textbf{282}, 727 (1987).

\bibitem{Colferai:2010wu}
  D.~Colferai, F.~Schwennsen, L.~Szymanowski, S.~Wallon,
  JHEP {\bf 1012} (2010) 026
  [arXiv:1002.1365 [hep-ph]].

\bibitem{Caporale:2012ih}
  F.~Caporale, D.Yu.~Ivanov, B.~Murdaca, A.~Papa,
  Nucl.\ Phys.\ B {\bf 877} (2013) 73
  [arXiv:1211.7225 [hep-ph]].

\bibitem{Ducloue:2013wmi}
  B.~Duclou\'e, L.~Szymanowski, S.~Wallon,
  JHEP {\bf 1305} (2013) 096
  [arXiv:1302.7012 [hep-ph]].

\bibitem{Ducloue:2013bva}
  B.~Duclou\'e, L.~Szymanowski, S.~Wallon,
  Phys.\ Rev.\ Lett.\  {\bf 112} (2014) 082003
  [arXiv:1309.3229 [hep-ph]].
  
\bibitem{Caporale:2013uva}
  F.~Caporale, B.~Murdaca, A.~Sabio Vera, C.~Salas,
  Nucl.\ Phys.\ B {\bf 875} (2013) 134
  [arXiv:1305.4620 [hep-ph]].

\bibitem{Ducloue:2014koa}
  B.~Duclou\'e, L.~Szymanowski, S.~Wallon,
  Phys.\ Lett.\ B {\bf 738} (2014) 311
  [arXiv:1407.6593 [hep-ph]].

\bibitem{Caporale:2014gpa}
  F.~Caporale, D.Yu.~Ivanov, B.~Murdaca, A.~Papa,
  Eur.\ Phys.\ J.\ C {\bf 74}, no. 10, 3084 (2014)
  [Eur.\ Phys.\ J.\ C {\bf 75}, no. 11, 535 (2015)]
  [arXiv:1407.8431 [hep-ph]].

\bibitem{Ducloue:2015jba}
  B.~Duclou\'e, L.~Szymanowski, S.~Wallon,
  Phys.\ Rev.\ D {\bf 92} (2015) no.7,  076002
  [arXiv:1507.04735 [hep-ph]].

\bibitem{Caporale:2015uva}
  F.~Caporale, D.Yu.~Ivanov, B.~Murdaca, A.~Papa,
  Phys.\ Rev.\ D {\bf 91} (2015) no.11,  114009
  [arXiv:1504.06471 [hep-ph]].

\bibitem{Celiberto:2015yba}
  F.G.~Celiberto, D.Yu.~Ivanov, B.~Murdaca, A.~Papa,
  Eur.\ Phys.\ J.\ C {\bf 75} (2015) no.6,  292
  [arXiv:1504.08233 [hep-ph]].

\bibitem{Celiberto:2015mpa}
  F.G.~Celiberto, D.Yu.~Ivanov, B.~Murdaca, A.~Papa,
  Acta Phys.\ Polon.\ Supp.\  {\bf 8} (2015) 935
  [arXiv:1510.01626 [hep-ph]].

\bibitem{Celiberto:2016ygs}
  F.G.~Celiberto, D.Yu.~Ivanov, B.~Murdaca, A.~Papa,
  Eur.\ Phys.\ J.\ C {\bf 76} (2016) no.4,  224
  [arXiv:1601.07847 [hep-ph]].

\bibitem{Celiberto:2016vva}
  F.G.~Celiberto, D.Yu.~Ivanov, B.~Murdaca, A.~Papa,
  PoS DIS {\bf 2016} (2016) 176
  [arXiv:1606.08892 [hep-ph]].

\bibitem{Caporale:2018qnm}
  F.~Caporale, F.G.~Celiberto, G.~Chachamis, D.~Gordo G{\'o}mez, A.~Sabio Vera,
  Nucl.\ Phys.\ B {\bf 935} (2018) 412
  [arXiv:1806.06309 [hep-ph]].

\bibitem{Chachamis:2015crx}
  G.~Chachamis,
  arXiv:1512.04430 [hep-ph].
  
\bibitem{Celiberto:2016hae}
  F.G.~Celiberto, D.Yu.~Ivanov, B.~Murdaca, A.~Papa,
  Phys.\ Rev.\ D {\bf 94} (2016) no.3,  034013
  [arXiv:1604.08013 [hep-ph]].

\bibitem{Celiberto:2016zgb}
  F.G.~Celiberto, D.Yu.~Ivanov, B.~Murdaca, A.~Papa,
  AIP Conf.\ Proc.\  {\bf 1819} (2017) no.1,  060005
  doi:10.1063/1.4977161
  [arXiv:1611.04811 [hep-ph]].

\bibitem{Celiberto:2017ptm}
  F.G.~Celiberto, D.Yu.~Ivanov, B.~Murdaca, A.~Papa,
  Eur.\ Phys.\ J.\ C {\bf 77} (2017) no.6,  382
  [arXiv:1701.05077 [hep-ph]].

\bibitem{Caporale:2015vya}
  F.~Caporale, G.~Chachamis, B.~Murdaca, A.~Sabio Vera,
  Phys.\ Rev.\ Lett.\  {\bf 116} (2016) no.1,  012001
  [arXiv:1508.07711 [hep-ph]].

\bibitem{Caporale:2015int}
  F.~Caporale, F.G.~Celiberto, G.~Chachamis, A.~Sabio Vera,
  Eur.\ Phys.\ J.\ C {\bf 76} (2016) no.3,  165
  [arXiv:1512.03364 [hep-ph]].

\bibitem{Caporale:2016soq}
  F.~Caporale, F.G.~Celiberto, G.~Chachamis, D.~Gordo~G{\'o}mez, A.~Sabio Vera,
  Nucl.\ Phys.\ B {\bf 910} (2016) 374
  [arXiv:1603.07785 [hep-ph]].

\bibitem{Caporale:2016vxt}
  F.~Caporale, F.G.~Celiberto, G.~Chachamis, A.~Sabio Vera,
  PoS DIS {\bf 2016} (2016) 177
  [arXiv:1610.01880 [hep-ph]].

\bibitem{Caporale:2016xku}
  F.~Caporale, F.G.~Celiberto, G.~Chachamis, D.~Gordo~G{\'o}mez, A.~Sabio Vera,
  Eur.\ Phys.\ J.\ C {\bf 77} (2017) no.1,  5
  arXiv:1606.00574 [hep-ph].

\bibitem{Celiberto:2016vhn}
  F.G.~Celiberto,
  Frascati Phys.\ Ser.\  {\bf 63} (2016) 43
  [arXiv:1606.07327 [hep-ph]].

\bibitem{Caporale:2016djm}
  F.~Caporale, F.G.~Celiberto, G.~Chachamis, D.~Gordo G\'omez, A.~Sabio Vera,
  AIP Conf.\ Proc.\  {\bf 1819} (2017) no.1,  060009
  [arXiv:1611.04813 [hep-ph]].

\bibitem{Caporale:2016lnh}
  F.~Caporale, F.G.~Celiberto, G.~Chachamis, D.~Gordo G\'omez, A.~Sabio Vera,
  EPJ Web Conf.\  {\bf 164} (2017) 07027
  [arXiv:1612.02771 [hep-ph]].

\bibitem{Caporale:2016zkc}
  F.~Caporale, F.G.~Celiberto, G.~Chachamis, D.~Gordo~G{\'o}mez, A.~Sabio Vera,
  Phys.\ Rev.\ D {\bf 95} (2017) no.7,  074007
  [arXiv:1612.05428 [hep-ph]].

\bibitem{Boussarie:2017oae}
  R.~Boussarie, B.~Duclou\'e, L.~Szymanowski, S.~Wallon,
  Phys.\ Rev.\ D {\bf 97} (2018) no.1,  014008
  [arXiv:1709.01380 [hep-ph]].

\bibitem{Bolognino:2018oth}
  A.D.~Bolognino, F.G.~Celiberto, D.Yu.~Ivanov, M.M.A.~Mohammed, A.~Papa,
  Eur.\ Phys.\ J.\ C {\bf 78} (2018) no.9,  772
  [arXiv:1808.05483 [hep-ph]].

\bibitem{Bolognino:2019yqj}
  A.D.~Bolognino, F.G.~Celiberto, D.Yu.~Ivanov, M.M.A.~Mohammed, A.~Papa,
  arXiv:1902.04511 [hep-ph].

\bibitem{Bolognino:2019cac}
  A.D.~Bolognino, F.G.~Celiberto, D.Yu.~Ivanov, M.M.A.~Mohammed, A.~Papa,
  arXiv:1906.11800 [hep-ph].

\bibitem{Motyka:2014lya}
  L.~Motyka, M.~Sadzikowski, T.~Stebel,
  JHEP {\bf 1505} (2015) 087
  [arXiv:1412.4675 [hep-ph]].

\bibitem{Brzeminski:2016lwh}
  D.~Brzeminski, L.~Motyka, M.~Sadzikowski, T.~Stebel,
  JHEP {\bf 1701} (2017) 005
  [arXiv:1611.04449 [hep-ph]].

\bibitem{Celiberto:2018muu}
  F.G.~Celiberto, D.~Gordo G{\'o}mez, A.~Sabio Vera,
  Phys.\ Lett.\ B {\bf 786} (2018) 201
  [arXiv:1808.09511 [hep-ph]].

\bibitem{Golec-Biernat:2018kem}
  K.~Golec-Biernat, L.~Motyka, T.~Stebel,
  JHEP {\bf 1812} (2018) 091
  [arXiv:1811.04361 [hep-ph]].

\bibitem{Deak:2018obv}
  M.~Deak, A.~van Hameren, H.~Jung, A.~Kusina, K.~Kutak, M.~Serino,
  Phys.\ Rev.\ D {\bf 99} (2019) no.9,  094011
  [arXiv:1809.03854 [hep-ph]].

\bibitem{Celiberto:2017nyx}
F.G.~Celiberto, D.Yu.~Ivanov, B.~Murdaca, A.~Papa,
Phys.\ Lett.\ B {\bf 777} (2018) 141
[arXiv:1709.10032 [hep-ph]].

\bibitem{Bolognino:2019ouc}
  A.D. Bolognino, F.G.~Celiberto, M.~Fucilla, D.Yu.~Ivanov, B.~Murdaca, A.~Papa,
  arXiv:1906.05940 [hep-ph].

\bibitem{Ahrens:2010zv}
  V.~Ahrens, A.~Ferroglia, M.~Neubert, B.D.~Pecjak, L.L.~Yang,
  JHEP {\bf 1009} (2010) 097
  [arXiv:1003.5827 [hep-ph]].

\bibitem{Fadin:1998fv}
  V.S.~Fadin, R.~Fiore,
  Phys.\ Lett.\ B {\bf 440} (1998) 359
  [hep-ph/9807472].

\bibitem{Fadin:2001dc}
  V.S.~Fadin, R.~Fiore,
  Phys.\ Rev.\ D {\bf 64} (2001) 114012
  [hep-ph/0107010].

\bibitem{Vera:2006un}
  A.~Sabio Vera,
  Nucl.\ Phys.\ B {\bf 746} (2006) 1
  [hep-ph/0602250].

\bibitem{Vera:2007kn}
  A.~Sabio Vera, F.~Schwennsen,
  Nucl.\ Phys.\ B {\bf 776} (2007) 170
  [hep-ph/0702158 [HEP-PH]].

\bibitem{Tanabashi:2018oca}
  M.~Tanabashi {\it et al.} [Particle Data Group],
  Phys.\ Rev.\ D {\bf 98} (2018) no.3,  030001.

\bibitem{Chatrchyan:2012dk}
  S.~Chatrchyan {\it et al.} [CMS Collaboration],
  JHEP {\bf 1204} (2012) 084
  [arXiv:1202.4617 [hep-ex]].

\bibitem{Chatrchyan:2012jua}
  S.~Chatrchyan {\it et al.} [CMS Collaboration],
  JINST {\bf 8} (2013) P04013
  [arXiv:1211.4462 [hep-ex]].

\bibitem{cernlib}
CERNLIB Homepage: \url{http://cernlib.web.cern.ch/cernlib}. 

\bibitem{Cuba:2005}
 T.~Hahn,
  Comput.\ Phys.\ Commun. {\bf 168} (2005) 78
  [arXiv:1408.6373 [hep-ph]].
  
\bibitem{ConcCuba:2015}
 T.~Hahn,
  J.\ Phys.\ Conf.\ Ser. {\bf 608} (2015) 1
  [arXiv:hep-ph/0404043].

\bibitem{VegasLepage:1978}
 G.P.~Lepage, 
 J.\ Comput.\ Phys. {\bf 27} (1978) 192.

\bibitem{Harland-Lang:2014zoa}
 L.A.~Harland-Lang, A.D.~Martin, P.~Motylinski, R.S.~Thorne,
 Eur.\ Phys.\ J.\ C {\bf 75} (2015) no.5,  204
 [arXiv:1412.3989 [hep-ph]].

\bibitem{Buckley:2014ana}
  A.~Buckley, J.~Ferrando, S.~Lloyd, K.~Nordstr{\"o}m, B.~Page, M.~R{\"u}fenacht, M.~SchÃ¶nherr, G.~Watt,
  Eur.\ Phys.\ J.\ C {\bf 75} (2015) 132
  [arXiv:1412.7420 [hep-ph]].

\end{thebibliography}
\end{document}